\documentclass[pre,twocolumn,showpacs,showkeys,preprintnumbers,amsmath,amssymb]{revtex4}

\usepackage{graphicx}
\usepackage{dcolumn}
\usepackage{bm}

\begin{document}


\title{Investigation of oscillation frequency and disorder induced dynamic phase transitions
in a quenched-bond diluted Ising ferromagnet}
\author{Erol Vatansever}
\author{\"{U}mit Ak{\i}nc{\i}}
\author{Yusuf Y\"{u}ksel}
\author{Hamza Polat}\email{hamza.polat@deu.edu.tr}

\affiliation{Department of Physics, Dokuz Eyl\"{u}l University, TR-35160 Izmir, Turkey}
\date{\today}
\begin{abstract}
Frequency evolutions of hysteresis loop area and hysteresis tools such as remanence and
coercivity of a  kinetic Ising model in the presence of quenched bond dilution are investigated
in detail. The kinetic equation describing  the time dependence  of the magnetization is derived
by means of effective-field theory with single-site correlations. It is found that the frequency
dispersions of hysteresis loop area, remanence and coercivity strongly depend on the  quenched
bond randomness, as well as applied field amplitude and oscillation frequency. In addition,
the shape of the hysteresis curves for a wide variety of Hamiltonian parameters is studied
and some interesting behaviors are found. Finally, a comparison of our  observations with those
of recently published studies is represented and it is shown that there exists a
qualitatively good agreement.
\end{abstract}

\pacs{64.60.Ht, 75.30.Kz, 75.60.-d, 75.70.Rf}
\keywords{Dynamic phase transitions, Quenched bond diluted kinetic Ising model,
Effective-field theory.} 
\maketitle

\section{Introduction}\label{intro}
When a ferromagnetic material is subject to a time dependent oscillating magnetic field, the system may
not respond to the external magnetic field instantaneously, and this situation leads to some interesting
phenomena. Namely, dynamic phase transitions (DPTs) and dynamic hysteresis behavior originate due to a
competition between time scales of the relaxation time of the system and oscillation period of the
external magnetic field. On the other hand, at high temperatures and for the high amplitudes of the
periodic magnetic field, the system is able to follow the external field with some delay while this
is not the case for low temperatures and small magnetic field amplitudes. This spontaneous symmetry
breaking indicates the presence of a DPT \cite{chakrabarti1_acharyya1} which shows itself in the
dynamic order parameter (DOP) which is defined as the time average of the magnetization over a full
period of the oscillating field. According to early experimental studies, dating back to last
century, there exists an empirical law for the hysteresis loop areal scaling \cite{steinmetz}.
Up to now, a fairly well numerous theoretical studies  have been dedicated to DPT and hysteresis
properties of kinetic Ising models by using various methods such as Monte Carlo (MC) simulations
\cite{lo, acharyya2_chakrabarti2, acharyya3, acharyya4, rao, sides1, sides2, sides3, korniss, zhu,
zhong,  acharyya6},  effective-field theory (EFT) \cite{shi, deviren} and mean-field theory (MFT)
\cite{tome, acharyya2_chakrabarti2, acharyya3, acharyya5, punya, wang, zimmer, luse, liu,sariyer}.
Apart from these, there exist a detailed geometrical description \cite{mayergoyz}, and also an
hysteresis criterion based on rate competition between the critical curvature and the
potential-barrier height \cite{gilmore}. By employing standard Metropolis MC algorithm with periodic
boundary conditions, the pure kinetic Ising model in a two dimensional square lattice has been
simulated by Lo and Pelcovits \cite{lo}, and  they found some evidences of a DPT. They also investigated
the behavior of dynamic hysteresis for  varying Hamiltonian parameters, however they did not intend to
make any classification for  dynamic hysteresis behavior. Afterwards, by making use of MC simulations
and MFT, Acharyya and Chakrabarti \cite{acharyya2_chakrabarti2} have presented a comprehensive
investigation of the pure kinetic Ising model and they showed that the hysteresis loops are asymmetric
in the dynamically ordered phase while in a dynamic disordered phase, they are symmetric around the
origin in a magnetization versus field amplitude plane. In Ref. \cite{acharyya3}, the attention has
been focused on the behavior of dynamic  hysteresis and dynamic loop area (DLA) of a pure kinetic Ising model
by using both MFT and MC calculations. In addition, the frequency variations of the coercive field
have been analyzed below the zero field critical temperatures. It has been concluded that the numerical
results for the dynamic coercive field show a power law frequency variation for both cases.
However, the coercive field calculated by MFT becomes frequency independent in the low frequency regime.
Differences between the two methods (MFT and MC) have been explained by using a kind of
Landau-type double well free energy mechanism. Moreover, there exist many previously published
studies such as the domain wall motion and nucleation theory \cite{nattermann1, nattermann2, misra, sch}
regarding the  hysteresis  loop mechanism \cite{lyuksyutov}.

On the experimental picture, it can be said that a great many of experimental studies have been devoted
to DPTs and hysteresis behaviors of Co films on a Cu (001) surface \cite{jiang}, [Co(4$\mathrm{A^{o}}$)/Pt(7$\mathrm{A^{o}}$)]
multi-layer system with strong perpendicular anisotropy \cite{robb}, thin polycrystalline Ni$_{80}$Fe$_{20}$ films \cite{choi},
ultrathin ferromagnetic Fe/Au(001) films \cite{he}, epitaxial Fe/GaAs(001) thin films \cite{lee},  epitaxial single ferromagnetic
fcc NiFe(001), fcc Co(001), and fcc NiFe/Cu/Co(001) layers \cite{lee2}, Fe$_{0.42}$Zn$_{0.58}$F$_{2}$ \cite{rivera}, finemet thin
films \cite{santi}, and epitaxial Fe/GaAs(001) and Fe/InAs(001) ultrathin films \cite{moore}. After some detailed experimental
investigations, it has been observed that experimental non-equilibrium dynamics of considered real magnetic systems strongly
resemble the dynamic behavior predicted from theoretical calculations of a kinetic Ising model. It is clear from these works
that there exists a strong evidence of qualitative consistency between theoretical and experimental studies.

Besides, the lattice models including impurities and defects have attracted considerable attention, because they are very
useful to investigate the behavior of disordered systems in nature. For this purpose, more effective models have been
introduced to examine the influences of disorder on the thermal and magnetic properties of real magnetic materials.
For example, the effects of impurities on driving-rate-dependent energy loss in a ferromagnet under the time dependent
magnetic field have been analyzed by Zheng and Li \cite{zheng_li} by using several well defined models  within the frameworks
of MFT and MC, and they found using MFT that, the hysteresis loop area is a power law function of the linear driving rate
as $A-A_{0}\propto h^{\beta}$, where, $A_{0}, h$ and  $\beta$ are the static hysteresis loop area, the linear driving rate
and scaling exponent of the system, respectively. Very recently,  the quenched site and bond diluted kinetic Ising models
under the influence of a time dependent oscillating magnetic field have been analyzed by making use of
EFT \cite{akinci, vatansever} on a two dimensional honeycomb lattice. In Ref. \cite{akinci},  the global phase
diagrams including the reentrant phase transitions are presented by the authors for site diluted kinetic Ising model,
and they  showed that the coexistence regions disappear for sufficiently weak dilution of lattice sites.
Following the same methodology, the authors have concentrated on the influences of quenched bond dilution
process on the dynamic behavior of the system. After some detailed analysis, it has been found that the
impurities in a bond diluted kinetic Ising model give rise to a number of interesting  and unusual phenomena such as
reentrant phenomena and the impurities to have a tendency to destruct the first-order transitions and the dynamic
tricritical point \cite{vatansever}. Furthermore,  it has also been shown that dynamically ordered phase regions
get expanded with decreasing amplitude which is more evident at low frequencies.

Very recently, frequency dependencies of the dynamic phase boundary (DPB)  and magnetic hysteresis of Ising model in
an oscillating magnetic field have been examined by Punya \emph{et al.} \cite{punya}. Based on MFT, they have
represented the phase  diagram in a magnetic field versus temperature plane for varying frequencies, in order to
show the influence of the oscillation frequency of the external field on the hysteretic response of the system.
According to their calculations, it has been found that the frequency dispersion of the DLA, the remanence and
coercivity can be categorized into three distinct types for a fixed temperature.
Although the detailed investigations and classifications of the hysteretic behaviors of pure kinetic Ising model have been
done within the MFT \cite{punya}, as far as we know, the dynamic hysteretic behavior within the EFT has not been  studied
for disordered kinetic Ising models. Actually, the main motivation for our work comes from the recent analysis of Ref. \cite{punya}
for the investigation of hysteresis behavior and dynamic phase diagrams of kinetic Ising model.
Therefore in this work, we intend to probe the effects of the quenched bond dilution process on the dynamic
hysteresis behavior of kinetic Ising model in the presence of a time-dependent oscillating external magnetic
field by using the EFT with correlations based on the exact Van der Waerden identity for a spin-1/2 system.
Present method is quite superior to MFT, since thermal fluctuations are partly considered in EFT.  The outline of the
paper can be summarized as follows: The dynamic equation of motion, DOP and DLA of the quenched bond diluted kinetic
Ising model have been introduced in the next section. The numerical results and related discussions are given in
Section \ref{results}, and finally Section \ref{conclude} contains our conclusions.

\section{Formulation}\label{formulation}
The  kinetic Ising model is given by the time dependent Hamiltonian
\begin{equation}\label{eq1}
\mathcal{H}=-\sum_{<ij>}J_{ij}S_{i}^{z}S_{j}^{z}-H(t)\sum_{i}S_{i}^{z},
\end{equation}
where the spin variables $S_{i}^{z}=\pm1$ are defined on a honeycomb lattice $(q=3)$ and the first sum
in Eq. (\ref{eq1}) is over the nearest neighbor pairs of spins. We assume that the nearest-neighbor
interactions are randomly diluted on the lattice according to the probability distribution function
\begin{equation}\label{eq2}
P(J_{ij})=p\delta(J_{ij}-J)+(1-p)\delta(J_{ij}),
\end{equation}
where $p$ denotes the concentration of active bonds. The term $H(t)$ in Eq. (\ref{eq1}) is a time
dependent external magnetic field which is defined as
\begin{equation}\label{eq3}
H(t)=h\cos(\omega t),
\end{equation}
where $h$ and $w=2\pi f$ represent the amplitude and the angular frequency of the oscillating
field, respectively. If the system evolves according to a Glauber-type stochastic process \cite{glauber}
at  a rate of $1/\tau$ which represents the transitions per unit time then the dynamic equation of motion
can be obtained as follows
\begin{equation}\label{eq4}
\tau\frac{d\langle S_{i}^{z}\rangle}{dt}=-\langle S_{i}^{z}\rangle+\left\langle\tanh
\left[\frac{E_{i}+H(t)}{k_{B}T}\right]\right\rangle,
\end{equation}
where $E_{i}=\sum_{j}J_{ij}S_{j}$ is the local field acting on the lattice site $i$, and $k_{B}$ and
$T$ denote the Boltzmann constant and temperature, respectively.
If we apply the differential operator technique \cite{honmura_kaneyoshi,kaneyoshi1}
in Eq. (\ref{eq4}) by taking into account the random configurational averages we get
\begin{equation}\label{eq5}
\frac{dm}{dt}=-m+\left\langle\left\langle \prod_{j=1}^{q=3} A_{ij}+S_{j}^{z}
B_{ij}\right\rangle\right\rangle_{r} f(x)|_{x=0},
\end{equation}
where $A_{ij}=\cosh(J_{ij}\nabla)$, $B_{ij}=\sinh(J_{ij}\nabla)$, and $m=\langle\langle S_{i}^{z}\rangle\rangle_{r}$
represents the average magnetization. $\nabla=\partial/\partial x$ is a differential operator, and the inner $\langle...\rangle$ and
the outer $\langle...\rangle_{r}$ brackets represent the thermal and configurational averages,
respectively. When the right-hand side of Eq. (\ref{eq5}) is expanded, the multispin correlation
functions appear. The simplest approximation, and one of the most frequently adopted is to decouple
these correlations according to
\begin{equation}\label{eq6}
\left\langle\left\langle
S_{i}^{z}S_{j}^{z}...S_{l}^{z}\right\rangle\right\rangle_{r}\cong\left\langle\left\langle
S_{i}^{z}\right\rangle\right\rangle_{r}\left\langle\left\langle
S_{j}^{z}\right\rangle\right\rangle_{r}...\left\langle\left\langle
S_{l}^{z}\right\rangle\right\rangle_{r},
\end{equation}
for $i\neq j \neq...\neq l$ \cite{tamura_kaneyoshi}. If we expand the right-hand side
of Eq. (\ref{eq5}) with the help of Eq. (\ref{eq6}) then we obtain the following dynamic
effective-field equation of motion for the magnetization of the quenched bond diluted kinetic Ising model
\begin{equation}\label{eq7}
\frac{dm}{dt}=-m+\sum_{i=0}^{q=3}\lambda_{i}m^{i},
\end{equation}
where the coefficients $\lambda_{i}$  can be easily calculated by employing the mathematical
relation $\exp(\alpha \nabla)f(x)=f(x+\alpha)$. Eq. (\ref{eq7}) describes the non-equilibrium
behavior of the system in the EFT formalism. Moreover, the time dependence of magnetization can
be one of two types according to whether they obey the following property or not
\begin{equation}\label{eq8}
m(t)=-m(t+\pi/\omega).
\end{equation}
A solution satisfying Eq. (\ref{eq8}) is called symmetric solution and it corresponds to a
paramagnetic or disordered phase. In this type of solution, the time dependent magnetization
oscillates around zero value. The second type of solutions which does not satisfy
Eq. (\ref{eq8}) is called non-symmetric solution which corresponds to ferromagnetic
(i.e. ordered) phase  where the time dependent magnetization oscillates around a non-zero value.

The time averaged magnetization over a full cycle of the oscillating magnetic field acts as
DOP which is defined as follows \cite{tome}
\begin{equation}\label{eq9}
Q=\frac{\omega}{2\pi}\oint m(t)dt,
\end{equation}
where $m(t)$ is a stable and periodic function. On the other hand, the DLA corresponding to
energy loss due to the hysteresis is defined as \cite{acharyya4}
\begin{equation}\label{eq10}
A=-\oint m(t)dH=h\omega\oint m(t)\sin(\omega t)dt.
\end{equation}
\noindent The behavior of DOP and DLA as a function of the temperature for selected Hamiltonian
parameters can be obtained by solving
Eqs. (\ref{eq7}), (\ref{eq9}) and (\ref{eq10})  numerically.
The remanent magnetization and coercivity  can be calculated by benefiting from hysteresis
loop ($m(t)-H(t)$ curve) in order to understand and clarify the behavior of the quenched bond diluted system.

\section{Results and Discussion}\label{results}
\begin{figure*}
\includegraphics[width=6.0cm]{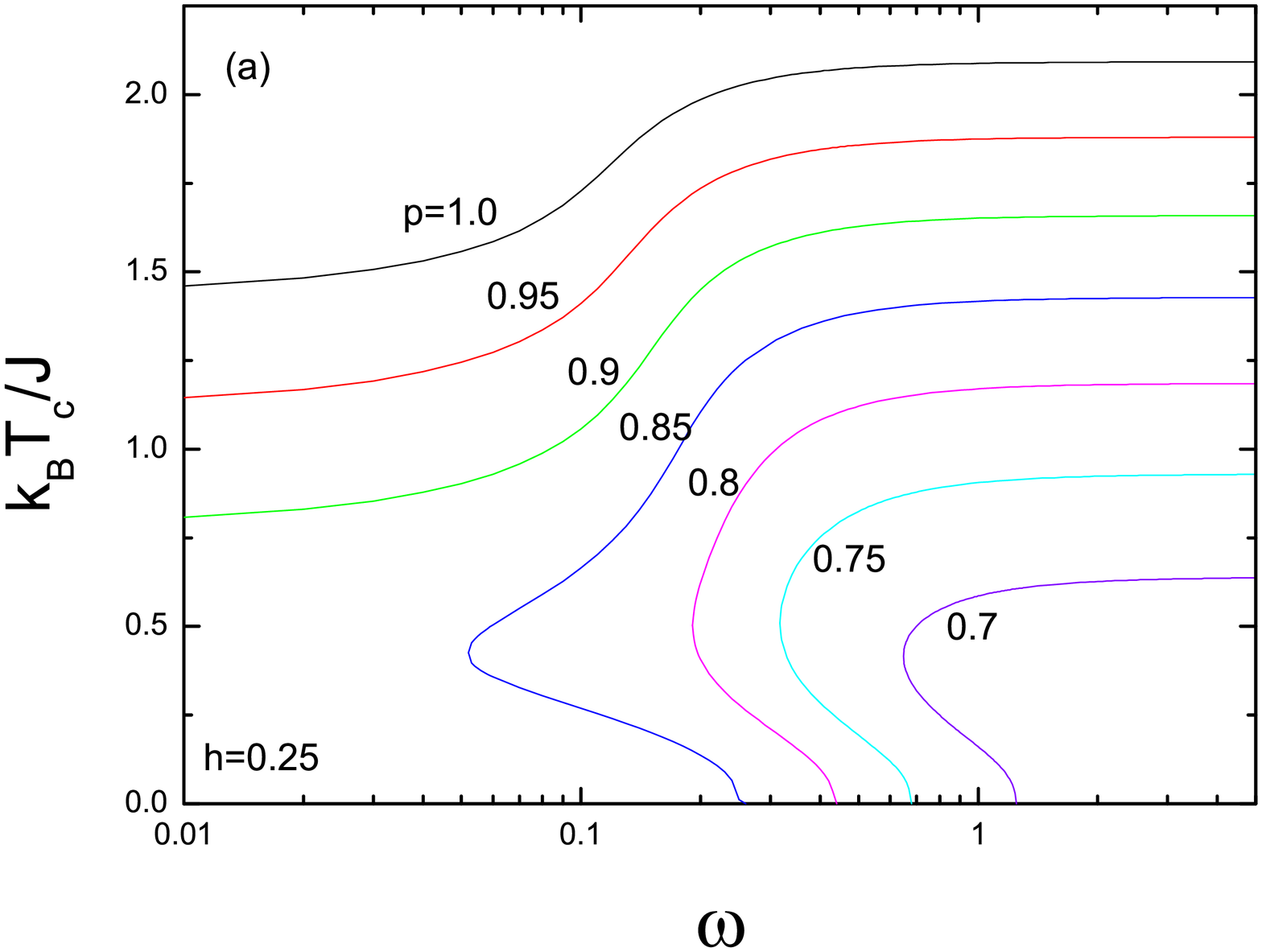}
\includegraphics[width=6.0cm]{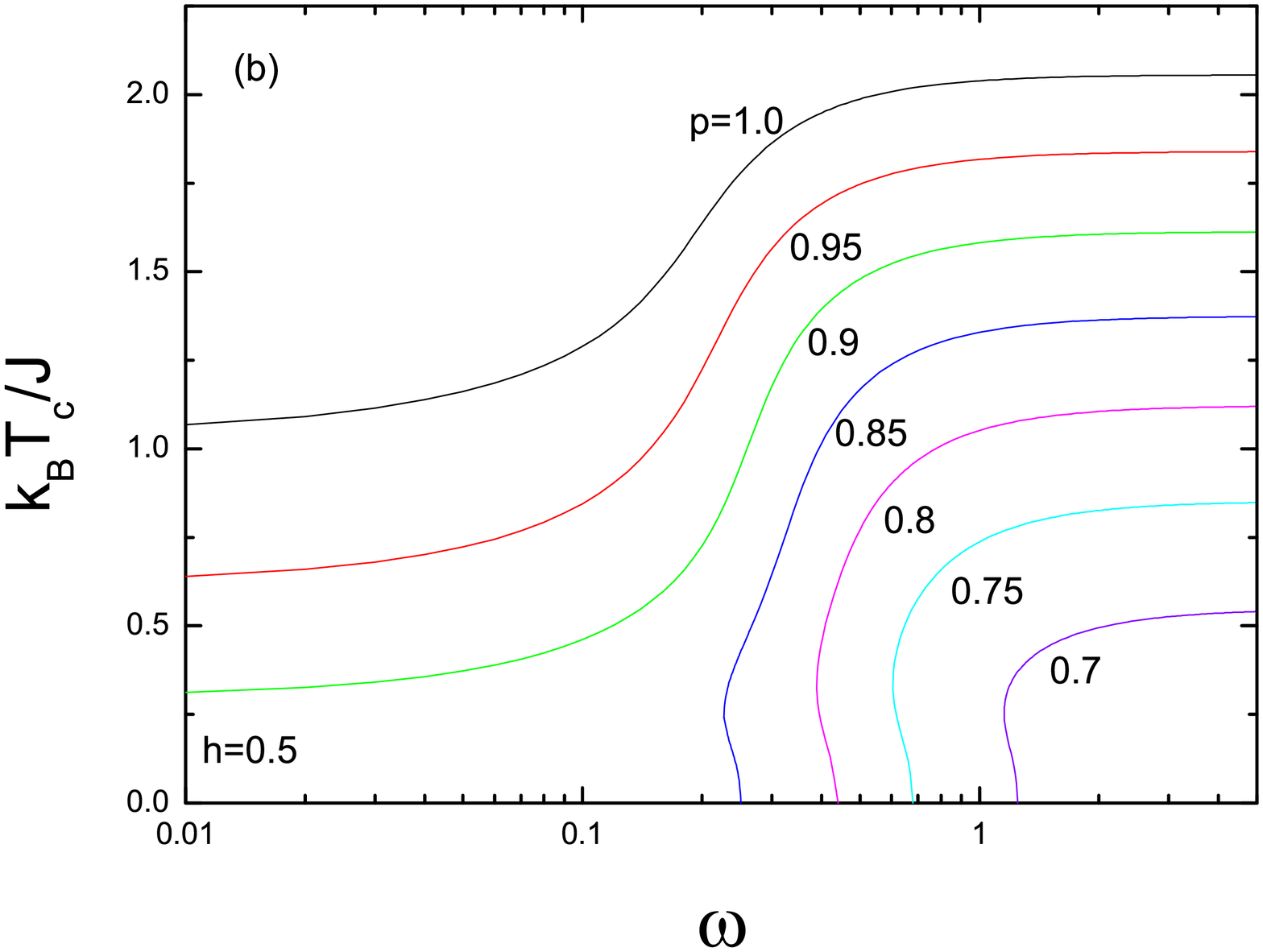}\\
\includegraphics[width=6.0cm]{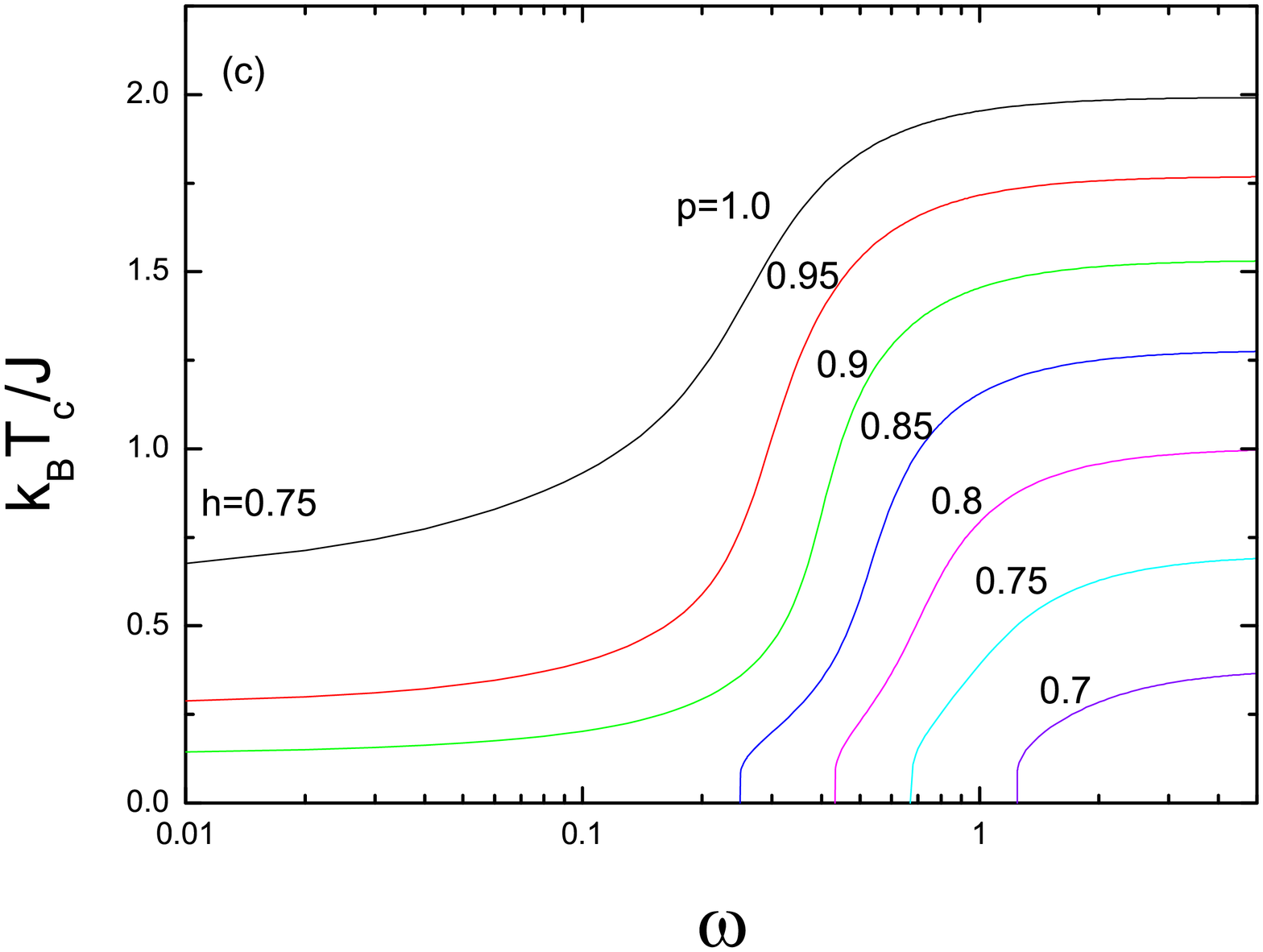}
\includegraphics[width=6.0cm]{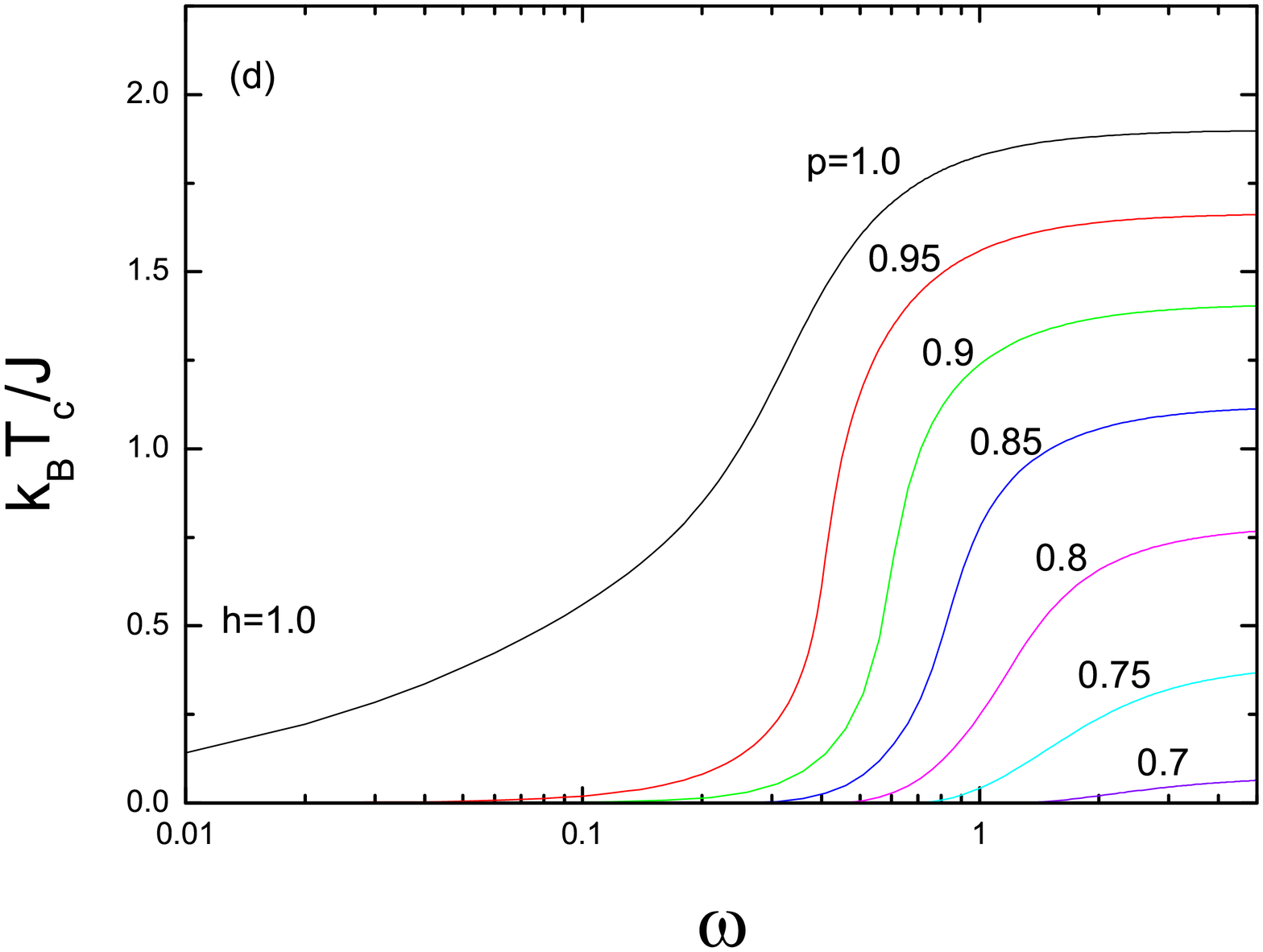}\\
\caption{Dynamic phase boundaries of the quenched bond diluted kinetic Ising system in a ($k_{B}T_{c}/J-\omega)$ plane with some selected
values of amplitude of the applied magnetic field. (a) $h=0.25$, (b) $h=0.50$, (c) $h=0.75$ and (d) $h=1.0$. The numbers
accompanying each curve denote the value of active bond concentrations.}\label{fig1}
\end{figure*}

In this section, in order to understand the dynamic evolution of the magnetic system in detail,
we will focus our attention on the hysteretic response and DPT properties of the quenched-bond
diluted kinetic Ising model under a time-dependent applied magnetic field. It is well
known that depending on the value of the applied field frequency,
the magnetic system can exist in a dynamically ordered or disordered phase.
From this point of view, we give the DPB in a $(k_{B}T_{c}/J-\omega)$ plane
with four selected amplitudes of the applied field $(h=0.25, 0.5, 0.75, 1.0)$, and for some
selected values of the active bond concentration in Figs. \ref{fig1}(a-d), in order to clarify
the influence of applied field frequency on the DPT properties. At first sight, one can easily
observe from  Figs. \ref{fig1}(a-d) that the dynamically ordered phases get expanded with increasing
applied field frequency for all considered active bond concentrations. Actually, this is an expected result since
as the frequency increases then the time-dependent magnetization can not follow applied magnetic field immediately,
hence the dynamically ordered phase region moves upward with the increasing frequency. We can also mention that
 aforementioned situations are also reported in previously published works \cite{chakrabarti1_acharyya1,
tome, deviren, punya, akinci, vatansever}.

On the other hand, as the value of $p$ decreases, starting from pure case (i.e. $p=1.0$), the magnetic system exhibits
unusual and interesting behaviors. For instance, dynamically disordered phases occur in the low temperature regions
and also reentrant behavior emerges. Another important consequence of decreasing $p$ is that the dynamically ordered
phases get narrower for all values of the considered applied field amplitudes. These observations can also be explained
by the following mechanism: The energy contribution which comes from the spin-spin interaction gets smaller when the
value of $p$  is decreased.  So, the system can undergo a DPT at a lower critical temperature, since the energy contribution
originating from the temperature and (or) magnetic field overcomes the ferromagnetic  spin-spin interactions. As a result of
this mechanism, the dynamically ordered phase regions in $(k_{B}T_{c}/J-\omega)$  plane get narrower. As discussed above,
the existence or absence of the reentrant behavior  strongly depends on the applied field amplitude, as well as oscillation frequency.
As an example, let us consider the curves with $p=0.85$ to show the influences of the applied field amplitude on the DPB, for selected
values of $h$. As seen in Figs. \ref{fig1}(a-d), for $h=0.25$ and $0.5$, the magnetic system exhibits reentrant behavior
whereas this behavior disappears with further increase in $h$, such as for $h=0.75$ and $1.0$.

\begin{figure*}
\center
\includegraphics[width=6.0cm]{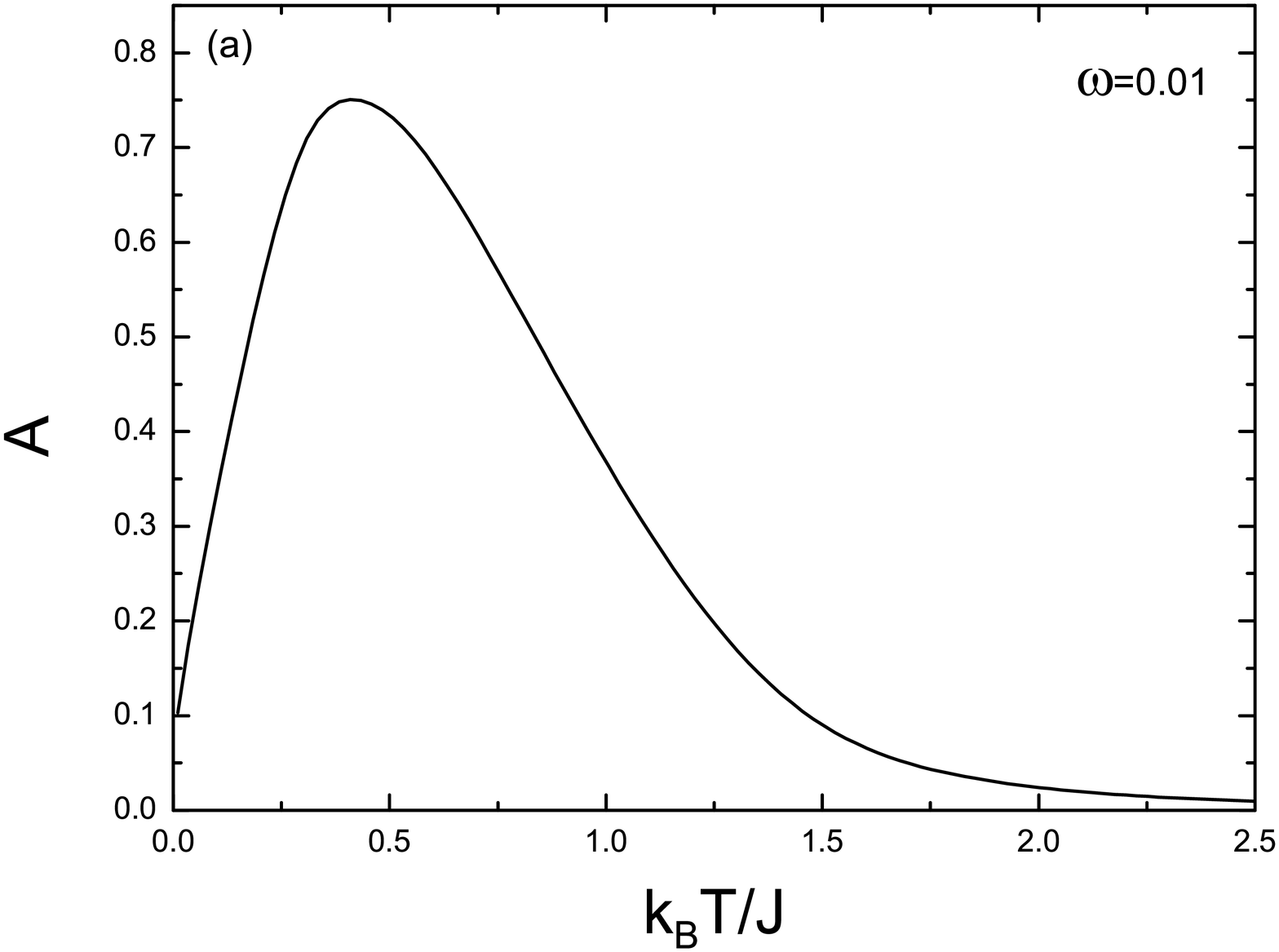}
\includegraphics[width=6.0cm]{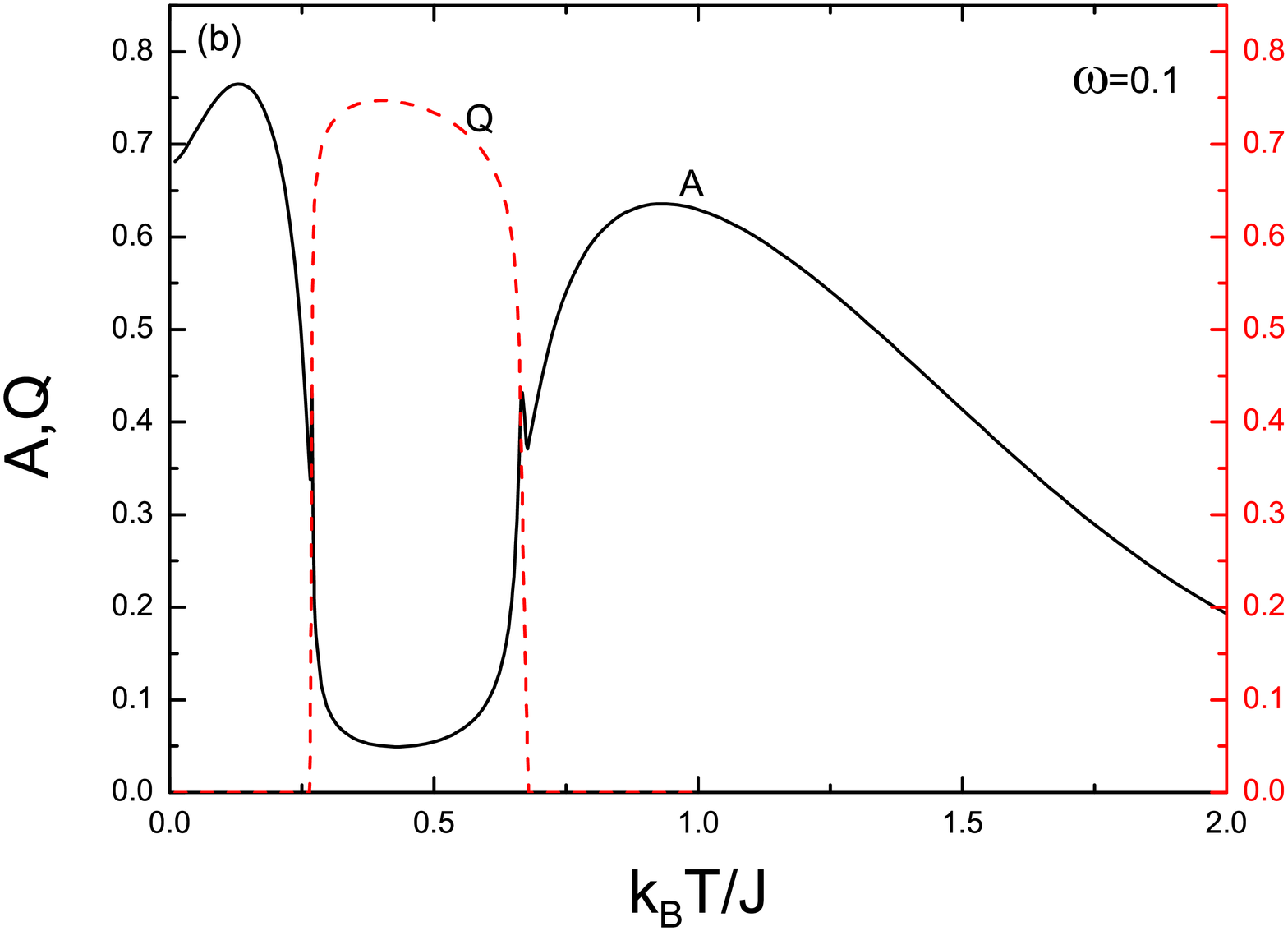}\\
\includegraphics[width=6.0cm]{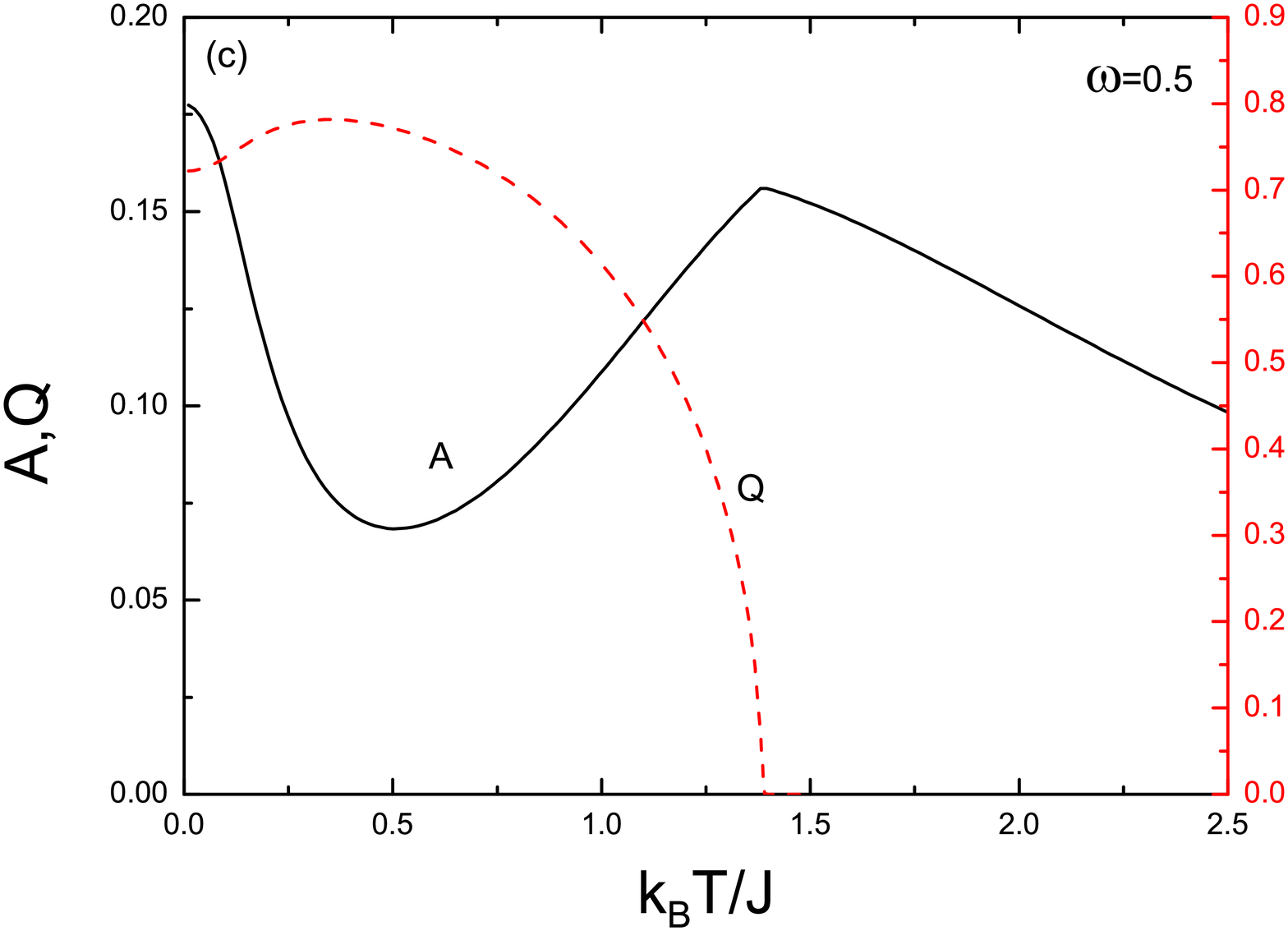}
\includegraphics[width=6.0cm]{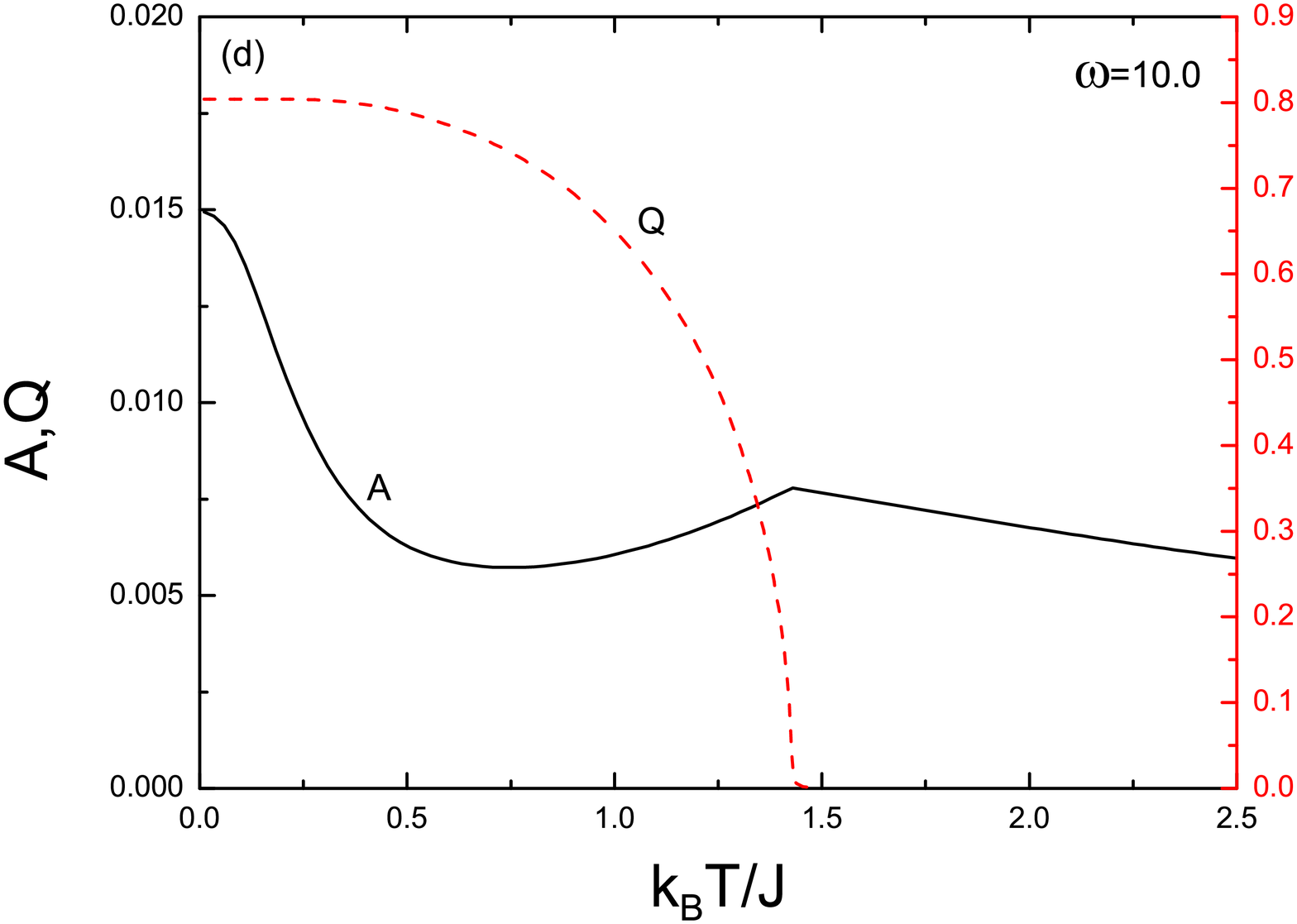}\\
\caption{Temperature variations of the dynamic order parameter and dynamic loop area
for considered values of the $h(=0.25)$ and $p(=0.85)$.}\label{fig2}
\end{figure*}

It is known that applied field frequencies naturally affect the dynamic evolution of a magnetic
system. In order to show the influences of the applied field frequencies on the system, the temperature
dependencies of the DOP \textbf{($Q$)} and DLA \textbf{($A$)} are depicted in Figs. \ref{fig2}(a-d)
for values of  $p=0.85$ and $h=0.25$ corresponding to Fig. \ref{fig1}(a). In accordance with this purpose,
we separated the frequency space into four regions: The order of the magnetic system
is dynamically paramagnetic in first region for $\omega=0.01$. One can see from Fig. \ref{fig2}(a) that
the DLA gradually increases and exhibits a local maxima with a Schottky-like rounded hump
and then tends to diminish with increasing temperature. On the other side, it is clear
from Fig. \ref{fig2}(b) that  there exists two successive second order DPTs for a selected
frequency value of $\omega=0.1$, and  this type of treatment corresponds to second region.
The first DPT is from paramagnetic  to ferromagnetic phase  while the second DPT is from ferromagnetic to
paramagnetic phase.  For the same Hamiltonian parameters, the DLA exhibits two local maxima and two whiskers and
the positions of the whiskers may correspond to DPT points. In Fig. \ref{fig2}(c), we represent the thermal variation
of the DOP and DLA curves for $\omega=0.5$ which indicates the third region. It can be easily
seen from Fig. \ref{fig2}(c) that  only one DPT emerges in magnetic system for $\omega=0.5$.
DOP curve for $\omega=0.5$ increases form its ground state value with increasing
temperature which originates from thermal agitations and it decreases with further increasing
temperatures and then falls to zero at a second order DPT temperature. Similarly, with the
increasing temperature, starting from zero, the DLA decreases and shows a local minima at the
dynamically  ordered phase region and then it exhibits a sharp peak near the DPT point. In the
following  analysis, let us consider the higher value of the applied field frequency such
as $\omega=10.0$ which refers to fourth region. As we can see from Fig. \ref{fig2}(d) that the magnetic system
exhibits a second order DPT. However, we should also mention that the value of phase transition temperature
observed for $\omega=10.0$ is slightly higher than that of $\omega=0.5$. Actually, this is an expected
result because an increase in the applied field frequency makes the occurrence of phase transition difficult,
hence the dynamically ordered phase gets expanded with the increasing applied field frequency. For given Hamiltonian
parameters, the DLA shows a peak above the second order DPT point. We should note that the existence of a
peak on the DLA near DPT point has also been indicated in previous investigations for various
models \cite{acharyya4,zhu, shi, deviren, huang}.

\begin{figure*}
\includegraphics[width=6.0cm]{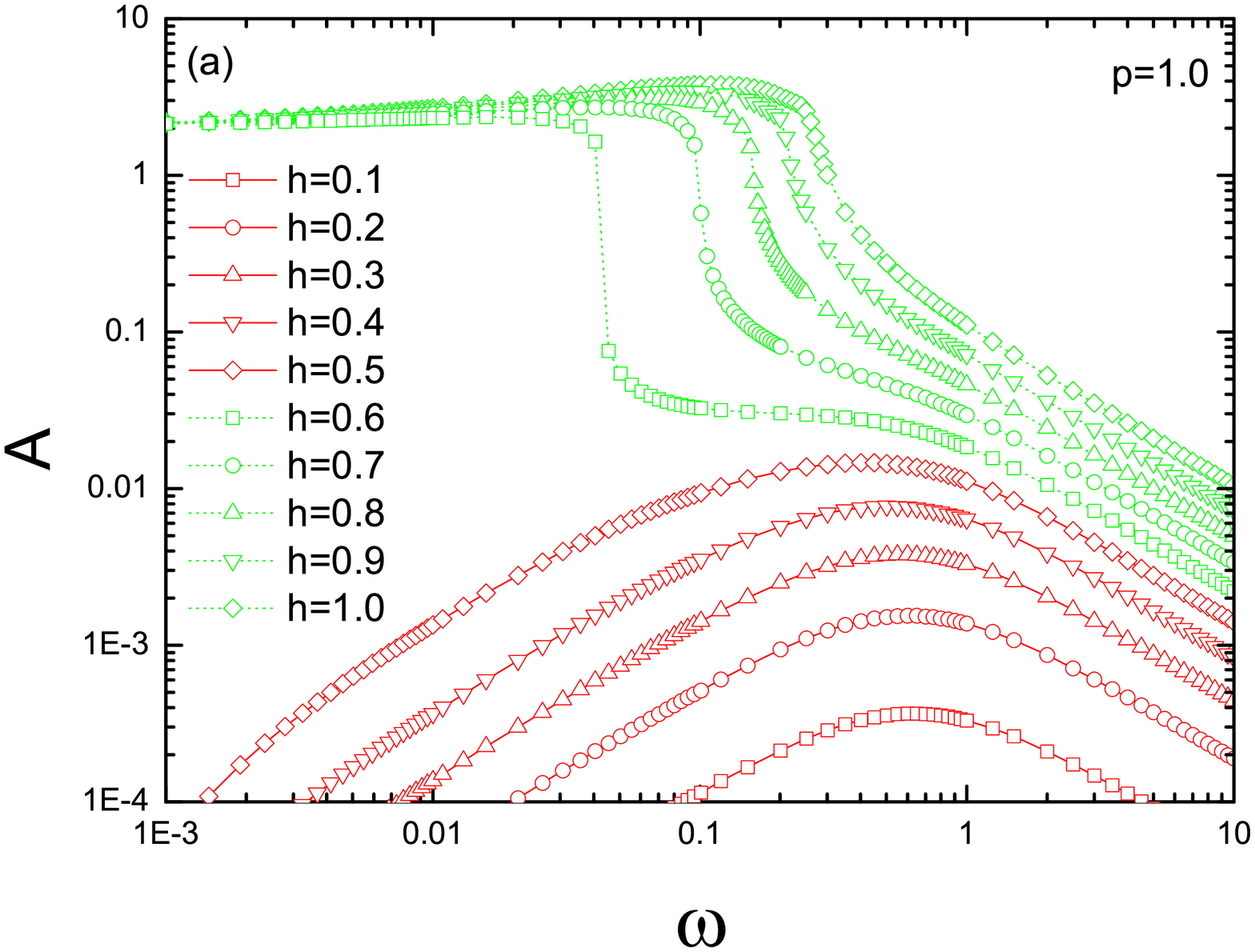}
\includegraphics[width=6.0cm]{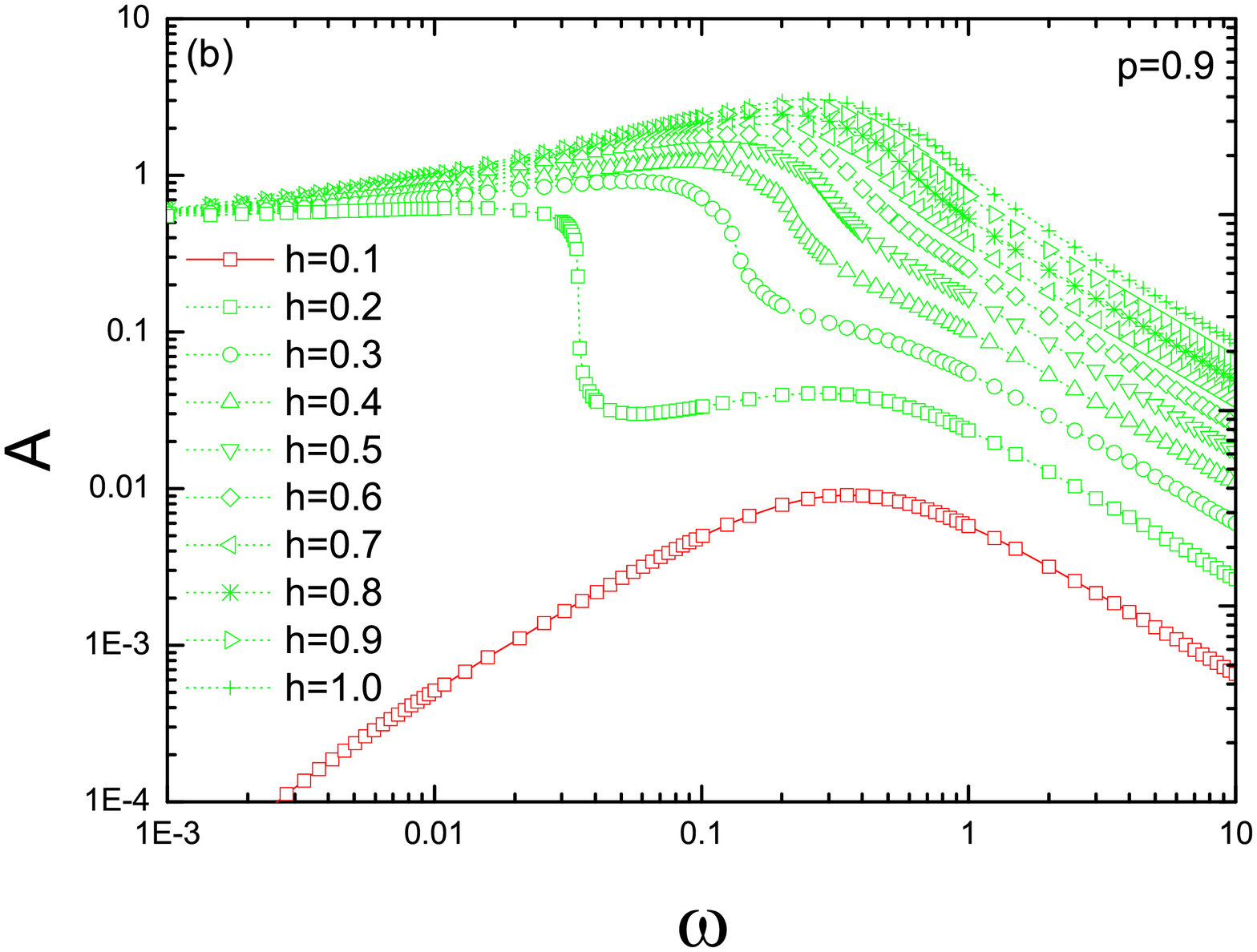}\\
\includegraphics[width=6.0cm]{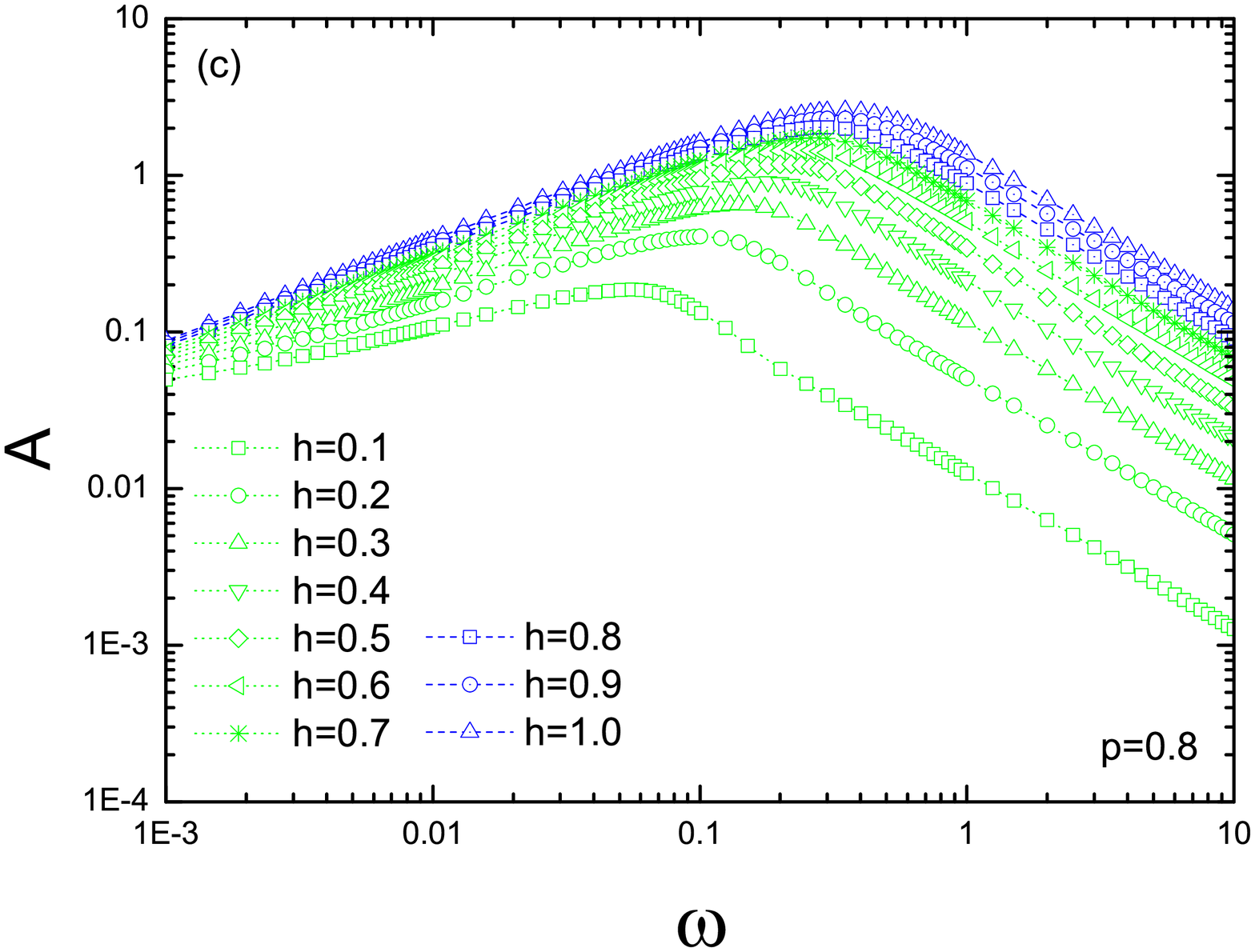}
\includegraphics[width=6.0cm]{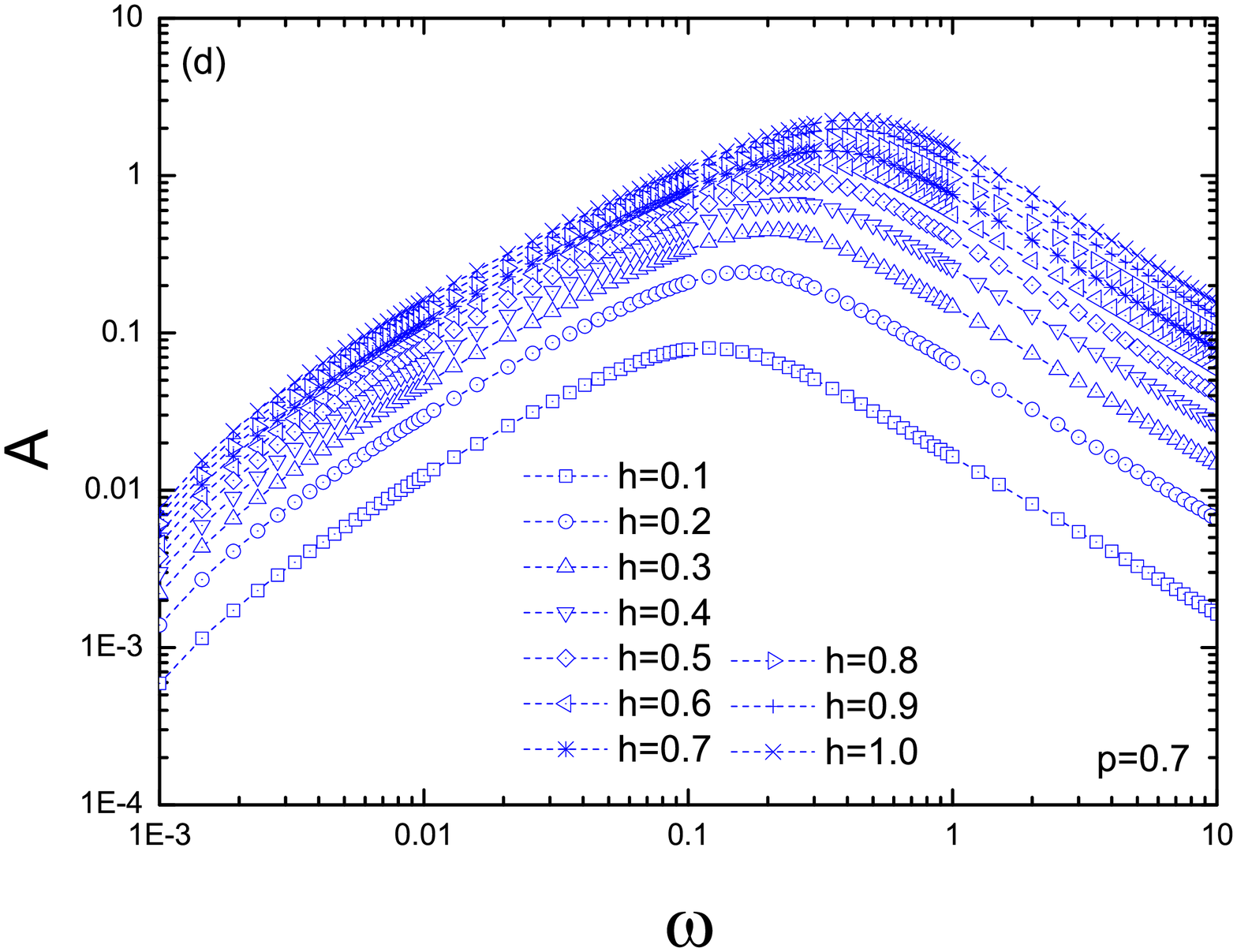}\\
\caption{The frequency dispersion of the hysteresis loop area curves with considered applied field amplitudes for selected values of
active bond concentrations (a) $p=1.0$, (b) $p=0.9$, (c) $p=0.8$ and (d) $p=0.7$.} \label{fig3}
\end{figure*}

We give the frequency dispersion curves of the DLA in Figs. \ref{fig3}(a-d) for four
selected $p$ values with a variety of $h$ values. We should note that the
reduced temperature in this study is fixed as $k_{B}T/J=1.0$ which is lower than the transition
temperature $T_{c}$ of  a pure system with EFT for $q=3$ in the absence of external field.
This selection of the temperature allows the system to undergo an amplitude, frequency or
disorder induced dynamic phase transition. Fig \ref{fig3}(a) is plotted for $p=1.0$ where all
ferromagnetic spin-spin exchange interactions $(J>0)$ are active. At first sight,
one can clearly observe in Fig. \ref{fig3}(a) that DLA curves can exhibit two types of
characteristic behavior. The first one corresponds to a dynamically  ordered phase where
the DLA increases as the applied field frequency increases and then exhibits a characteristic
rounded peak (or local maxima) for the relatively low field amplitude values such as $h=0.1, 0.2, 0.3, 0.4$
and $0.5$. These curves are identical to type-I curves of Ref. \cite{punya}. As is expected,
the DLA is relatively close to zero at the low and high applied field frequency regions
in the dynamically ordered phases. On the other hand, for $h=0.6, 0.7, 0.8, 0.9$ and
$1.0$ values, DLA curves start from a nonzero value and increases with increasing
frequency values then exhibit a finite cusp at a second order phase transition temperature
at which the shape of hysteresis loop changes from symmetric to asymmetric and also a
rounded hump in high frequency regime which corresponds to second type of characteristic
behavior. These curves correspond to type-II curves in Ref. \cite{punya}. Finite cusps are
found to be discontinuous for low $h$ and weak disorder. The rounded humps observed in DLA
curves in high frequency regime disappears for increasing $h$ and (or) decreasing $p$ values.
For instance, for $h=0.6$ with $p=1.0$, the DLA exhibits a discontinuous cusp and a rounded
distinct hump  while it displays only a continuous cusp for $h=1.0$ and $p=1.0$. We can mention
that the characteristic cusp of $(h,p)=(0.6,1.0)$ curve arises from the frequency evolution of
symmetric hysteresis loop, whereas the rounded hump of this curve originates from the frequency
evolution of asymmetric hysteresis loop. On the other side, it is known that the phase transition
temperature  decreases  with increasing amount of disorder. The effects of the aforementioned
situation on the frequency dispersion can be seen in Figs. \ref{fig3}(b-d) which have
been plotted for $p=0.9, 0.8,$ and $0.7$. One can observe in Figs. \ref{fig3}(b-d) that
as the disorder effects become dominant in the system then the representative type-I curves tend
to evolve into type-II, and type-II curves have a tendency to turn into type-III curves defined
in Ref. \cite{punya} which represent the paramagnetic DLA dispersion curves. In contrast to type-I curves,
type-III curves have a finite residual value in the limit of $\omega\rightarrow 0$. In addition, the
type-III curves are characterized by a rounded hump, as type-I curves. Based on these observations,
we see that DLA dispersion curves clearly reveal the signs of the oscillation frequency and disorder
induced dynamic phase transitions in the present system.

\begin{figure*}
\includegraphics[width=6.0cm]{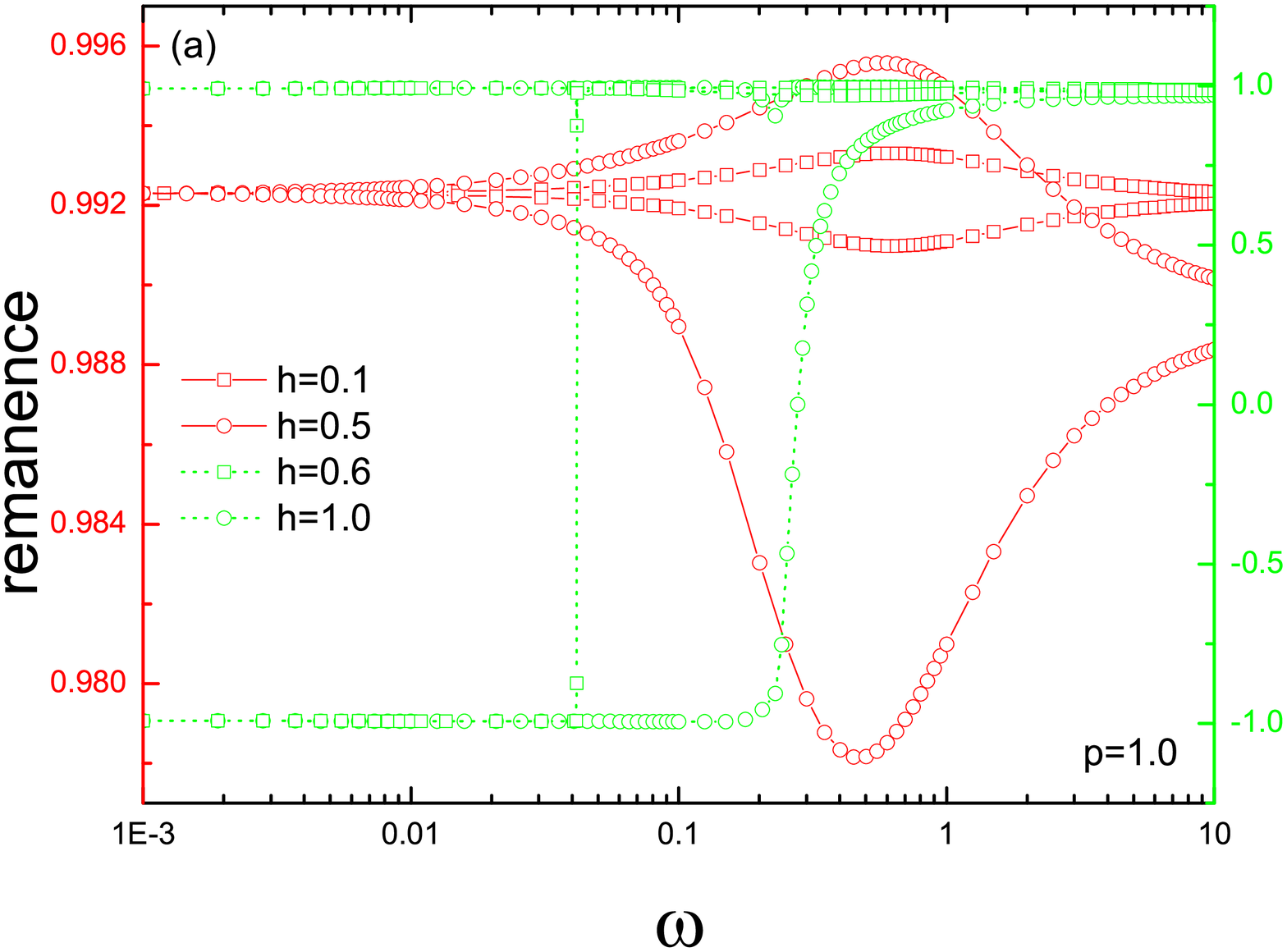}
\includegraphics[width=6.0cm]{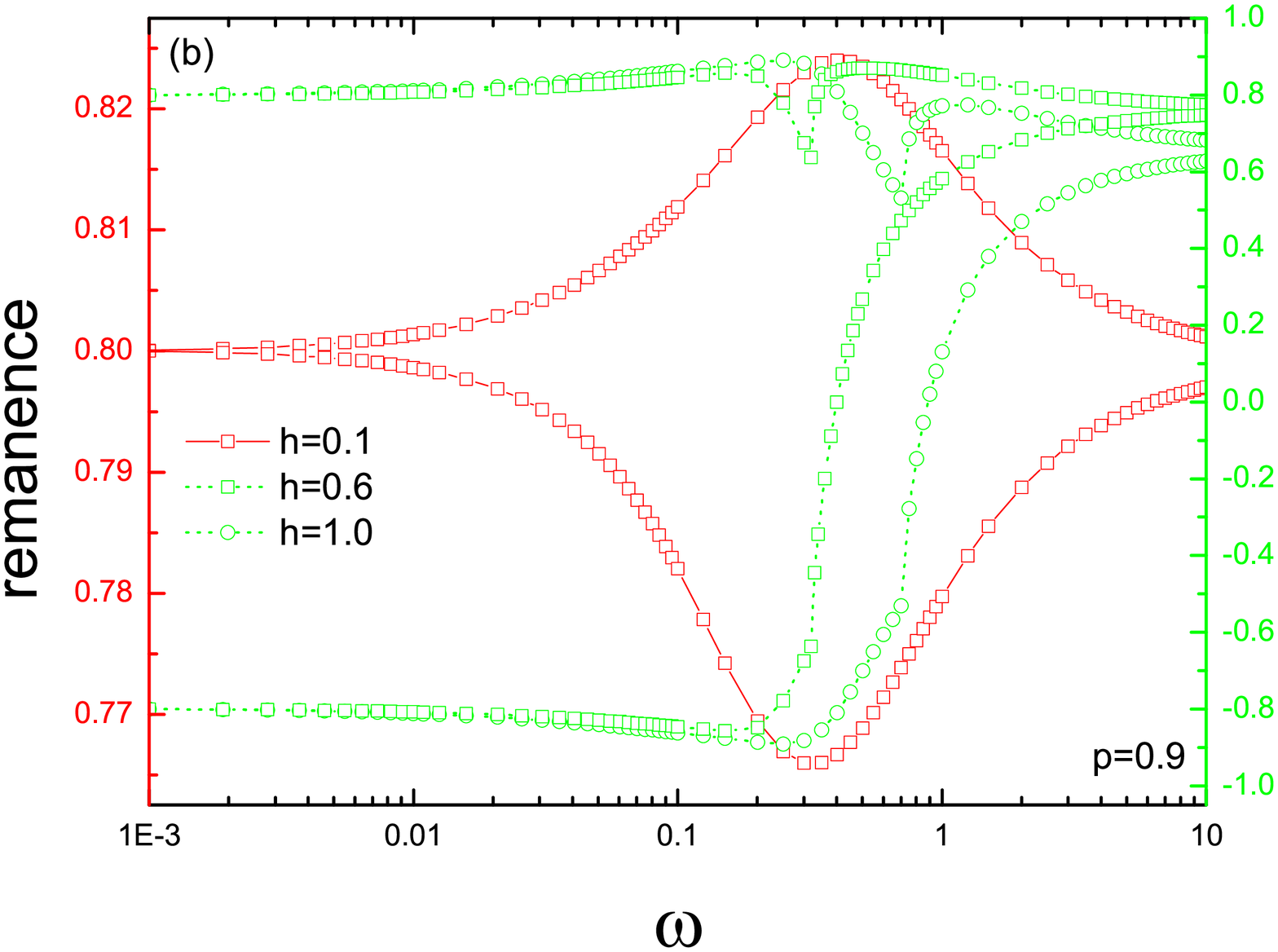}\\
\includegraphics[width=6.0cm]{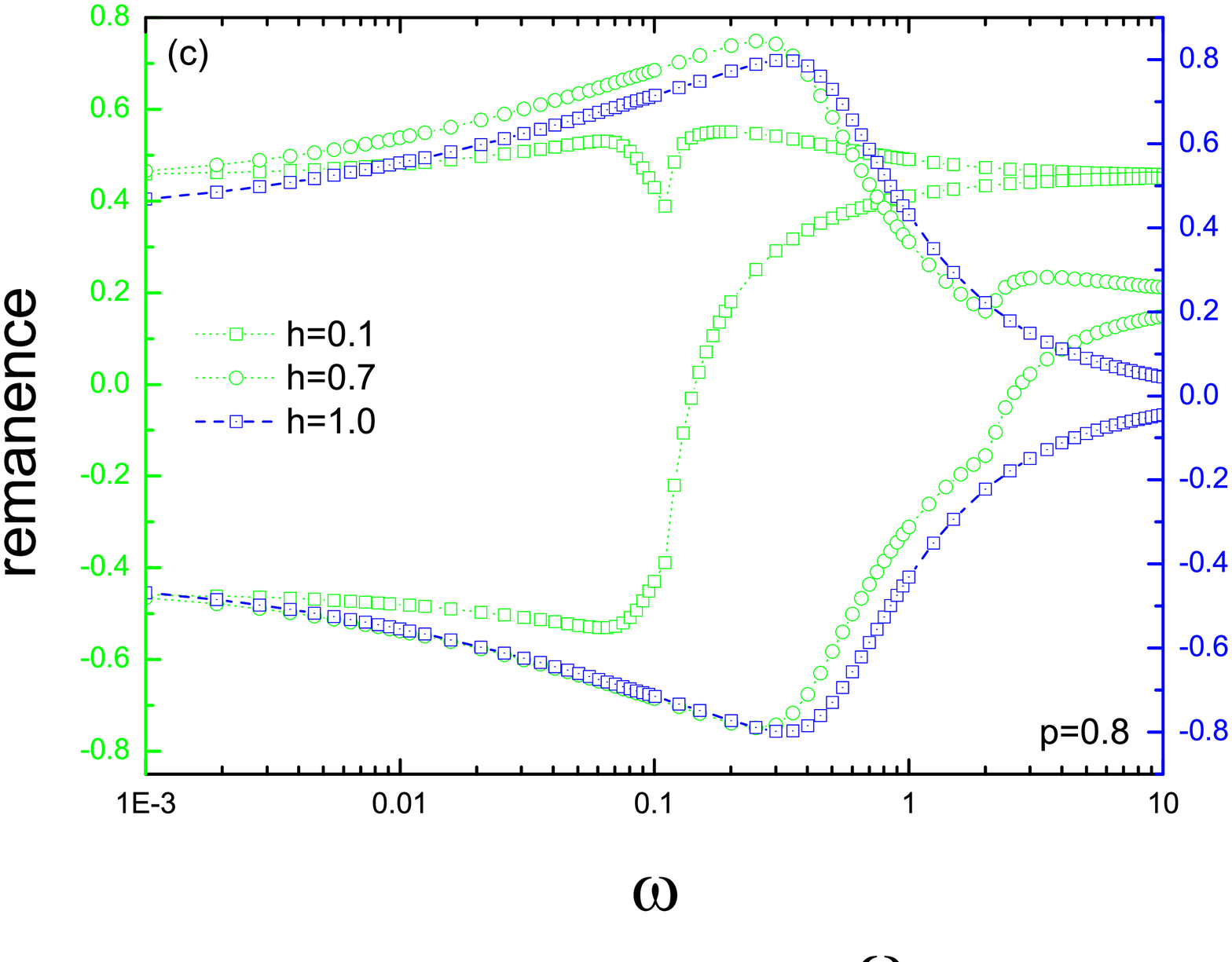}
\includegraphics[width=6.0cm]{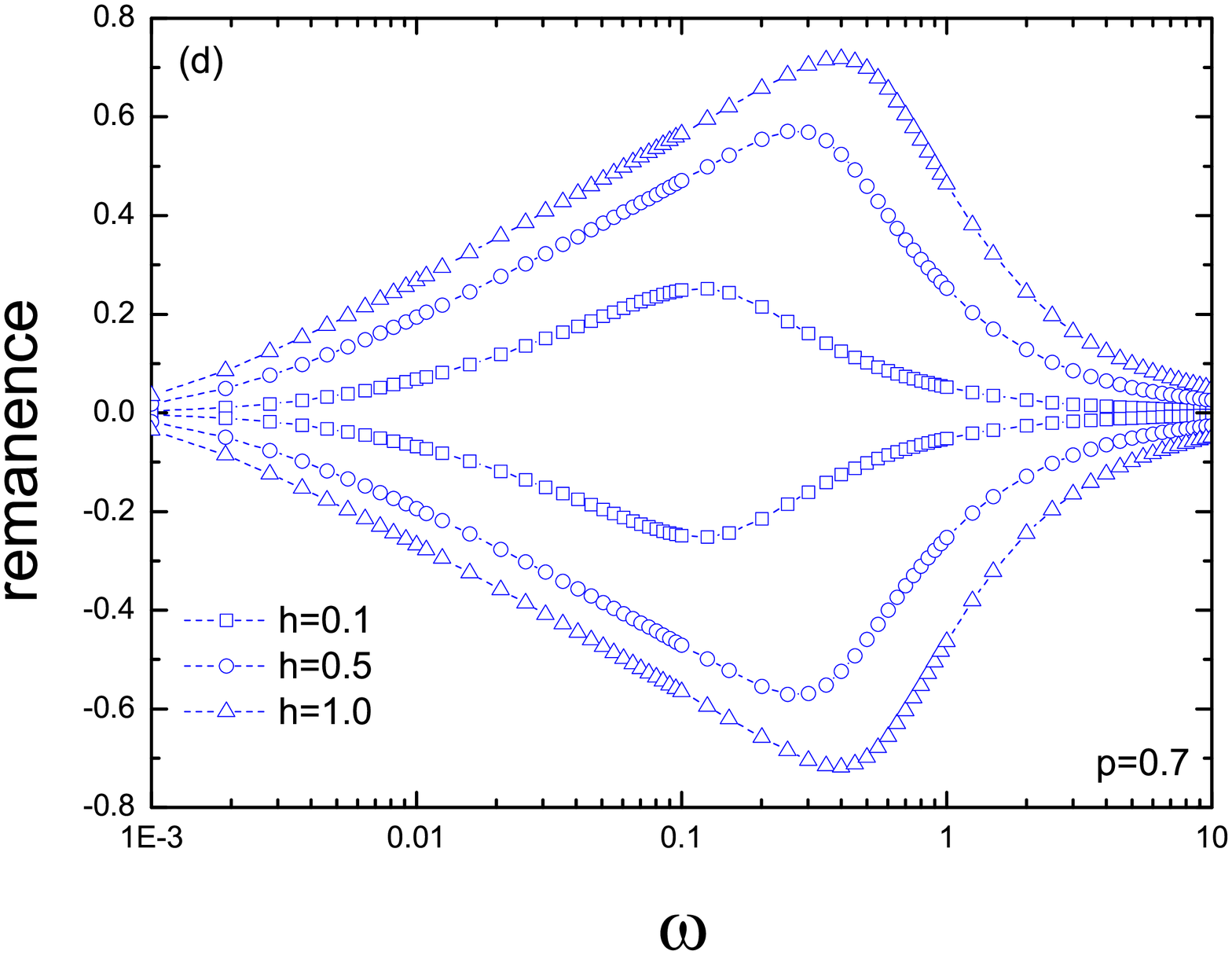}\\
\caption{Frequency dispersion of remanent magnetization for a variety of applied field amplitudes with  selected active bond
concentrations (a) $p=1.0$, (b) $p=0.9$, (c) $p=0.8$ and (d) $p=0.7$.}\label{fig4}
\end{figure*}

In Figs. \ref{fig4}(a-d), we present the frequency dependence of remanent magnetization
curves corresponding to DLA curves in Fig. \ref{fig3}. Here we select four typical values of disorder
parameter $p$ with a variety of field amplitude $h$ values. Remanent magnetization is defined as the residual
magnetization of a ferromagnetic material after an external magnetic field is removed.
Hence the remanence data can be easily extracted from calculated hysteresis loops.
At first glance, it can be easily seen that depending on amplitude and frequency of the external magnetic field,
as well as the amount of disorder, remanence curves can exhibit three distinct characteristics corresponding
to type-I, type-II and type-III profiles of DLA curves. Corresponding to type-I DLA curves, either branches of
remanent magnetization curves are positively valued which means that the system is ordered ferromagnetically in the
whole range of oscillation frequency $\omega$. Increasing the amount of disorder or amplitude of the external
field, representative type-I remanence curves may evolve to type-II and type-III curves in order.
Type-II remanence curves exhibit equally valued branches with opposite signs in the low frequency
region indicating that the system is in a dynamically paramagnetic state. As the frequency increases then the
negative branch of a representative type-II remanence curve changes its sign at a critical value of frequency,
hence we understand that the corresponding hysteresis loop loses its symmetry and becomes dynamically ordered.
Hence this behavior can be considered as a frequency induced phase transition. For sufficiently high amplitudes
and (or) for a sufficiently large amount of disorder, the observed type-II curves evolve to type-III characteristics.
In this case, either branches of remanence curves are symmetric with opposite signs which indicates that the system
is dynamically disordered for the whole range of oscillation frequency spectrum. Remanent magnetization is a vital
property for materials used in magnetic information storage devices. Namely, in case of some fixed values of external
field amplitude and oscillation frequency, as well as an amount of existing disorder, for a device made of a material
with higher remanence property there will be less chance that the information stored in the media will be erased due
to some external sources \cite{altavilla}. From this point of view, it is important to note that according to Fig. \ref{fig4}(d),
there exists an optimum field frequency for a disordered material under an oscillating magnetic field at which the
type-III remanence (i.e. the distance between positive and negative remanent magnetizations) exhibit a maximum value,
and as the field amplitude increases then the optimum frequency value shifts towards higher frequency values.

\begin{figure*}
\includegraphics[width=6.0cm]{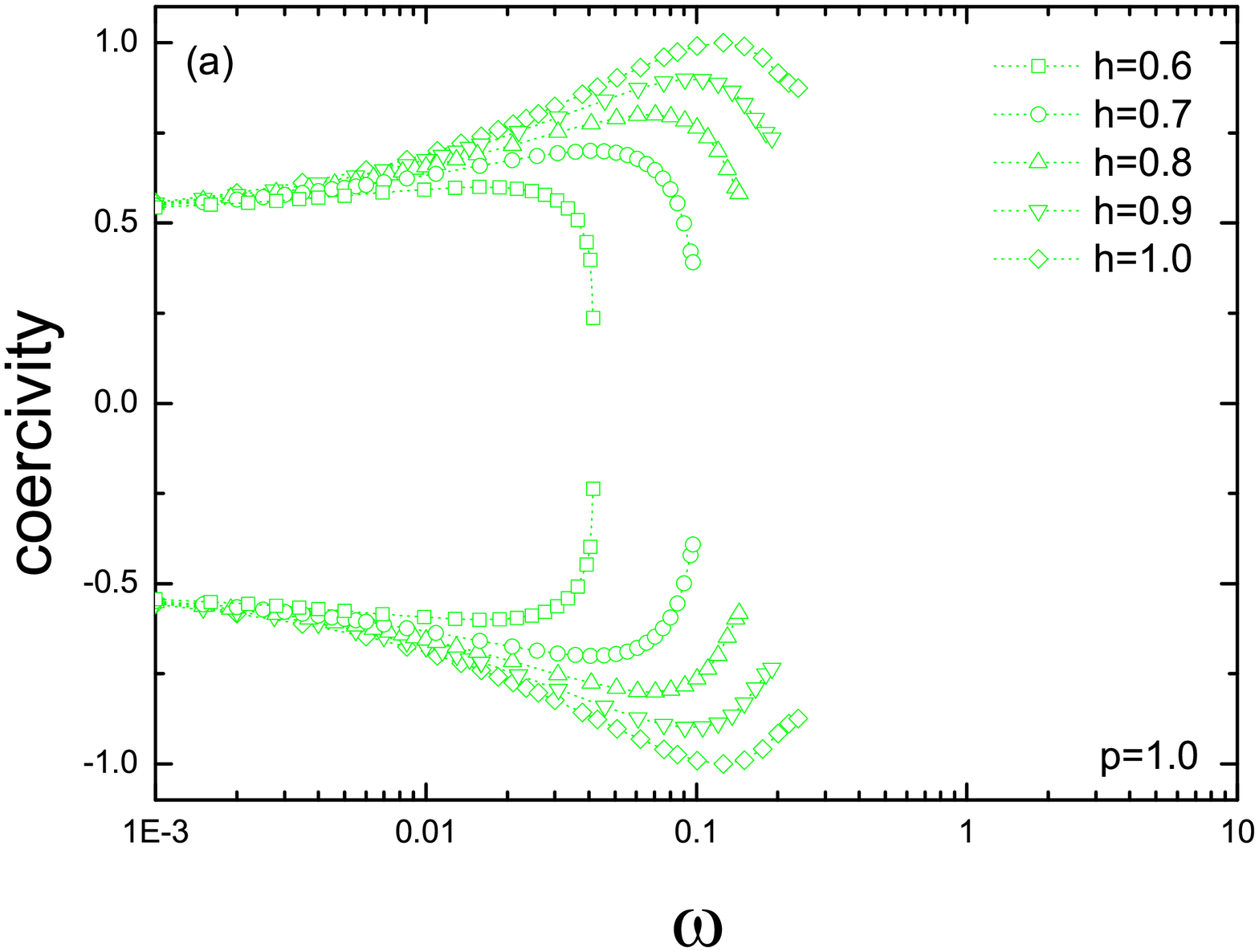}
\includegraphics[width=6.0cm]{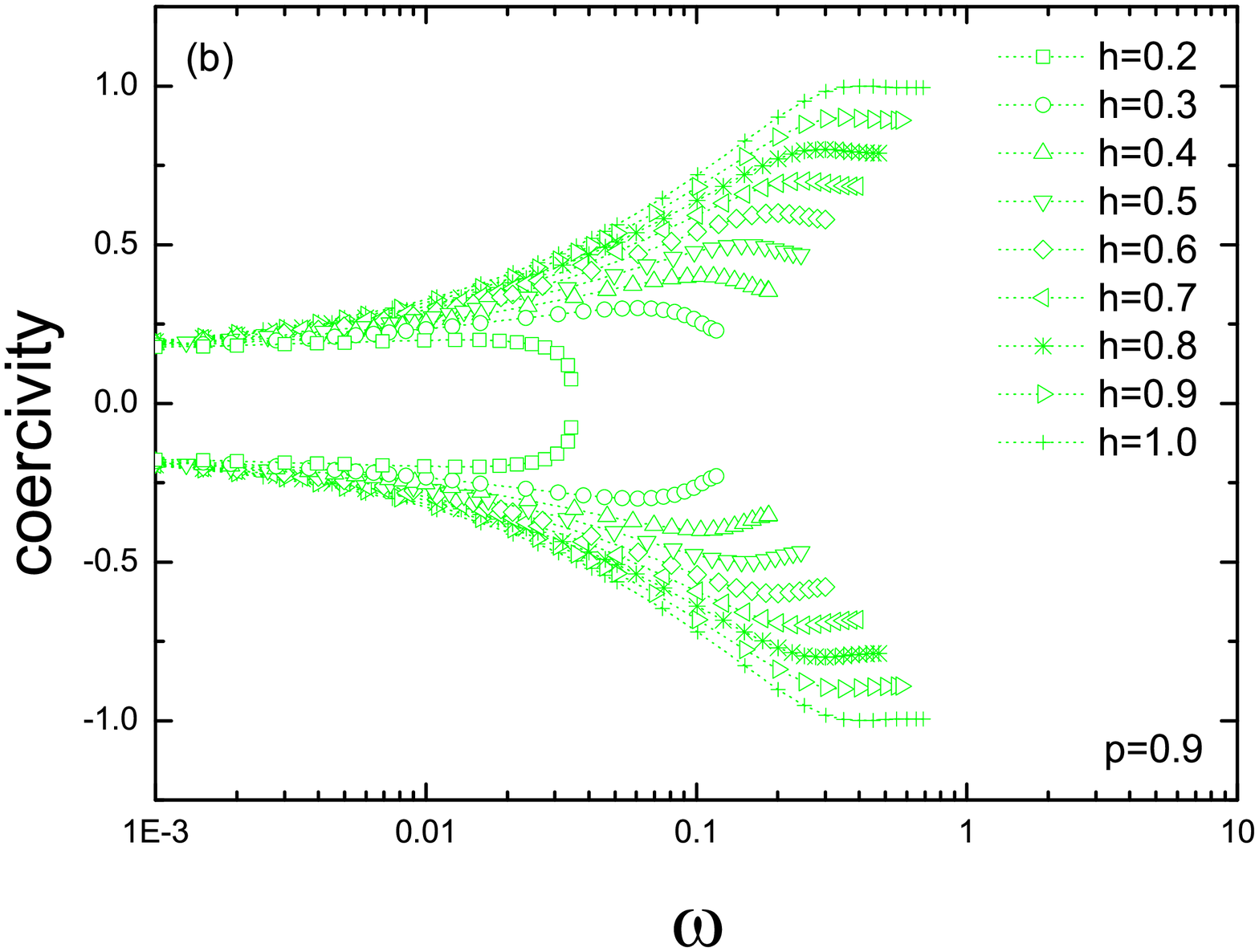}\\
\includegraphics[width=6.0cm]{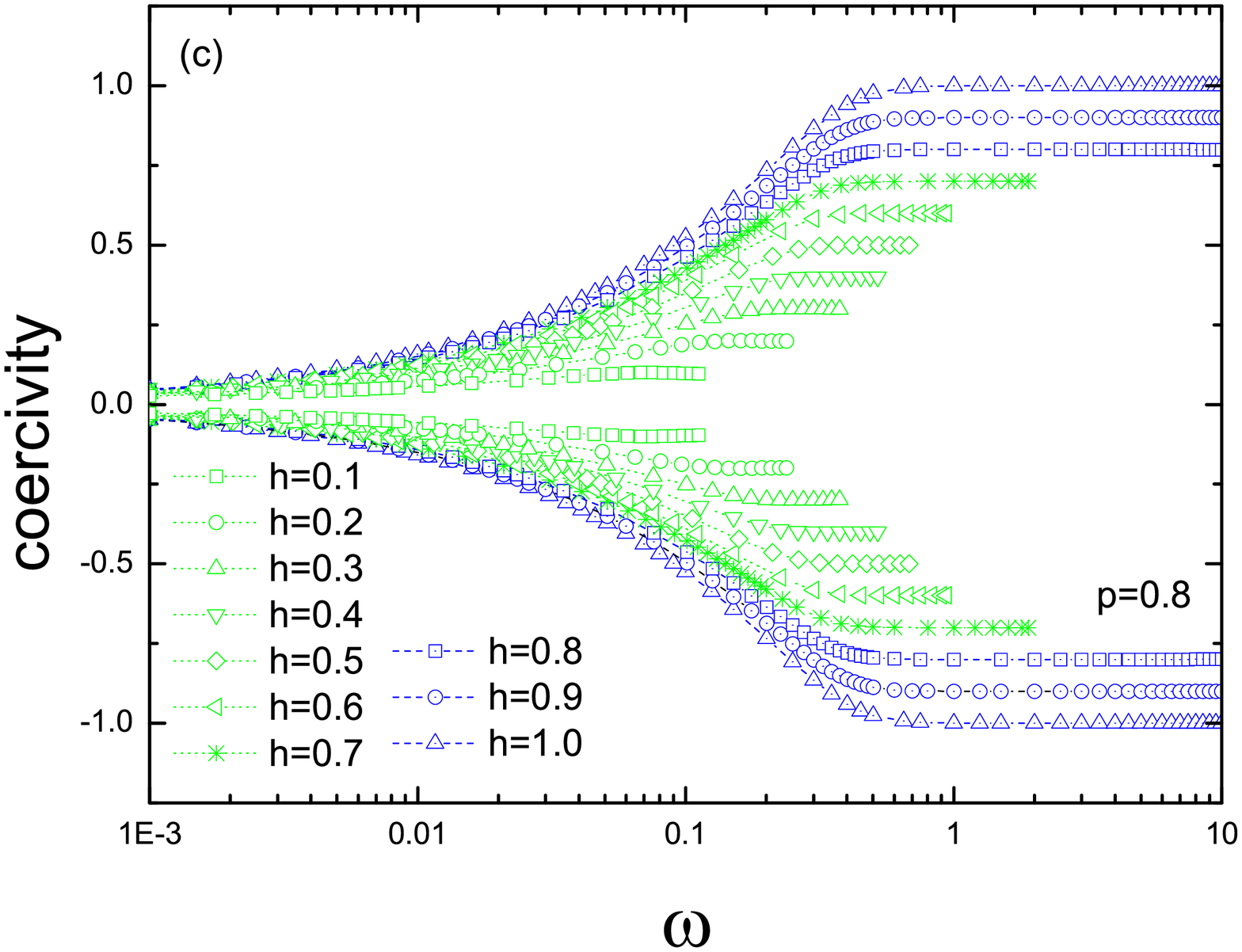}
\includegraphics[width=6.0cm]{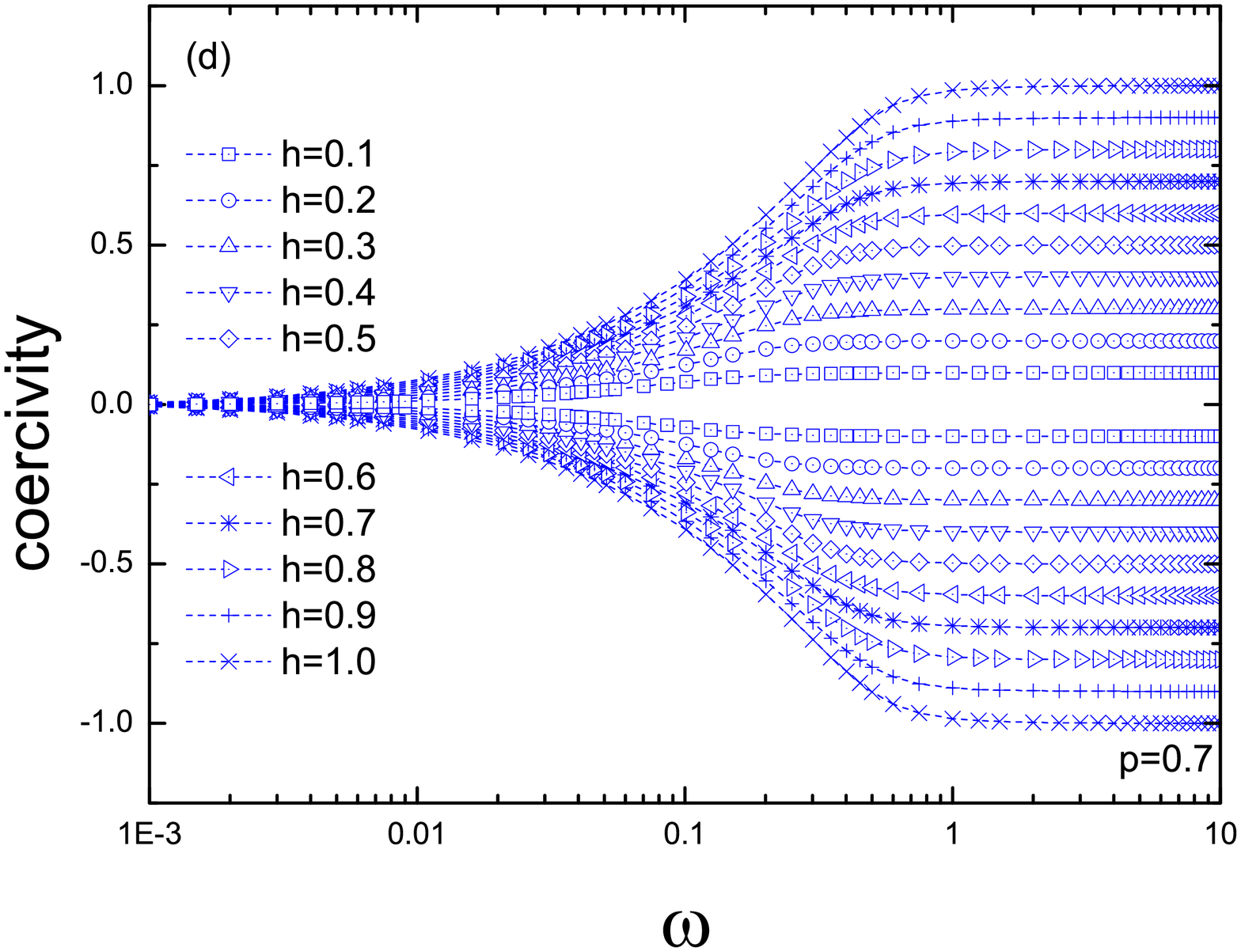}\\
\caption{Frequency dispersion of the coercivity for a variety of applied field amplitudes with selected  values
of  active bond concentrations (a) $p=1.0$, (b) $p=0.9$, (c) $p=0.8$ and (d) $p=0.7$.}\label{fig5}
\end{figure*}

As another characteristic feature of hysteresis loops, we have examined the coercivity of
the system which is defined as the minimum required reversal magnetic field, in order to reduce the
remanent magnetization to zero. For storage media, a material with a high coercivity is desired to
preserve the stored data from unwanted external factors. We represent the variation of the coercive
field curves  as a function of oscillation frequency in Figs. \ref{fig5}(a-d) for some selected
values of disorder parameter $p$ and external field amplitude $h$. For the DLA curves with type-I characteristics,
there is no coercivity for any field amplitude, oscillation frequency, or disorder, because all hysteresis
curves corresponding to type-I behavior have asymmetric shape. This means that the instantaneous magnetization
can not follow the external magnetic field. For type-II characteristics however, we see that at low frequencies,
absolute values of either branches in a coercivity curve of the system increase for a while then tend to decrease
with further increasing frequencies. After a critical frequency value, these branches lose their symmetric shape
then coercive fields disappear and a dynamic phase transition from a dynamic paramagnetic phase to a ferromagnetic
one takes place. We note that nonsymmetric coercivity data have not been shown in Fig. \ref{fig5}. We again see that
an oscillation frequency induced dynamic phase transition takes place in representative type-II curves.
If the magnetic field amplitude is large enough or if there exists a sufficiently strong disorder (see Figs. \ref{fig5}c and d),
coercivity curves may evolve from type-II to type-III characteristics. Since the type-III hysteresis curves are symmetric,
coercive field can be observed for the whole range of oscillation frequency values. However, for type-III behavior,
coercivity is rather small at low frequencies. As the field frequency increases then coercivity of the system
increases and in the limit of $\omega\rightarrow\infty$ it saturates to its maximum value. The maximum value of
coercivity is noting but just the applied amplitude of the external magnetic field for sufficiently high
frequencies. Hence, we can conclude that the greater amplitude of the external magnetic field, we get more
enhanced coercivity.

\begin{figure}
\includegraphics[width=6.0cm]{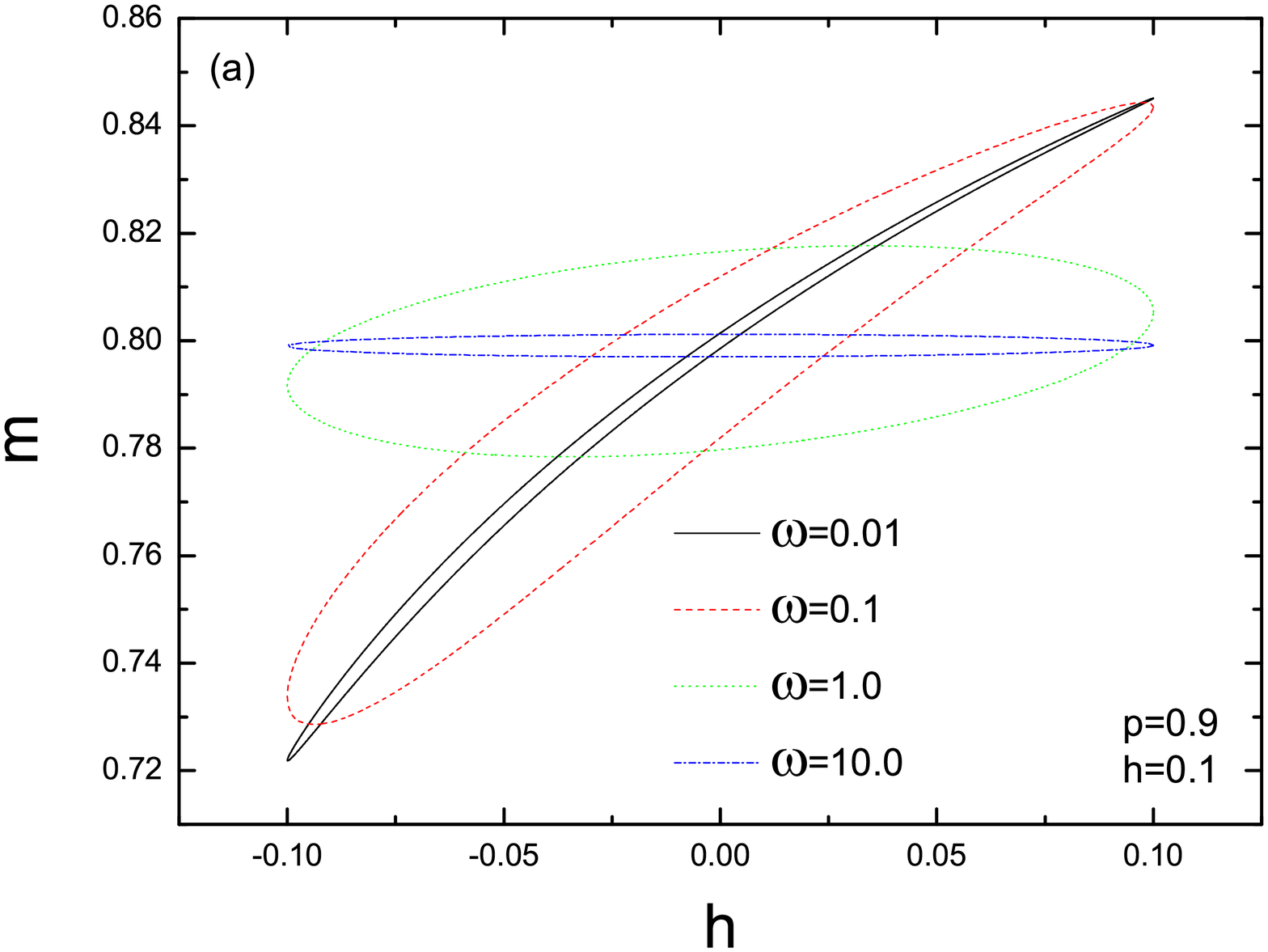}
\includegraphics[width=6.0cm]{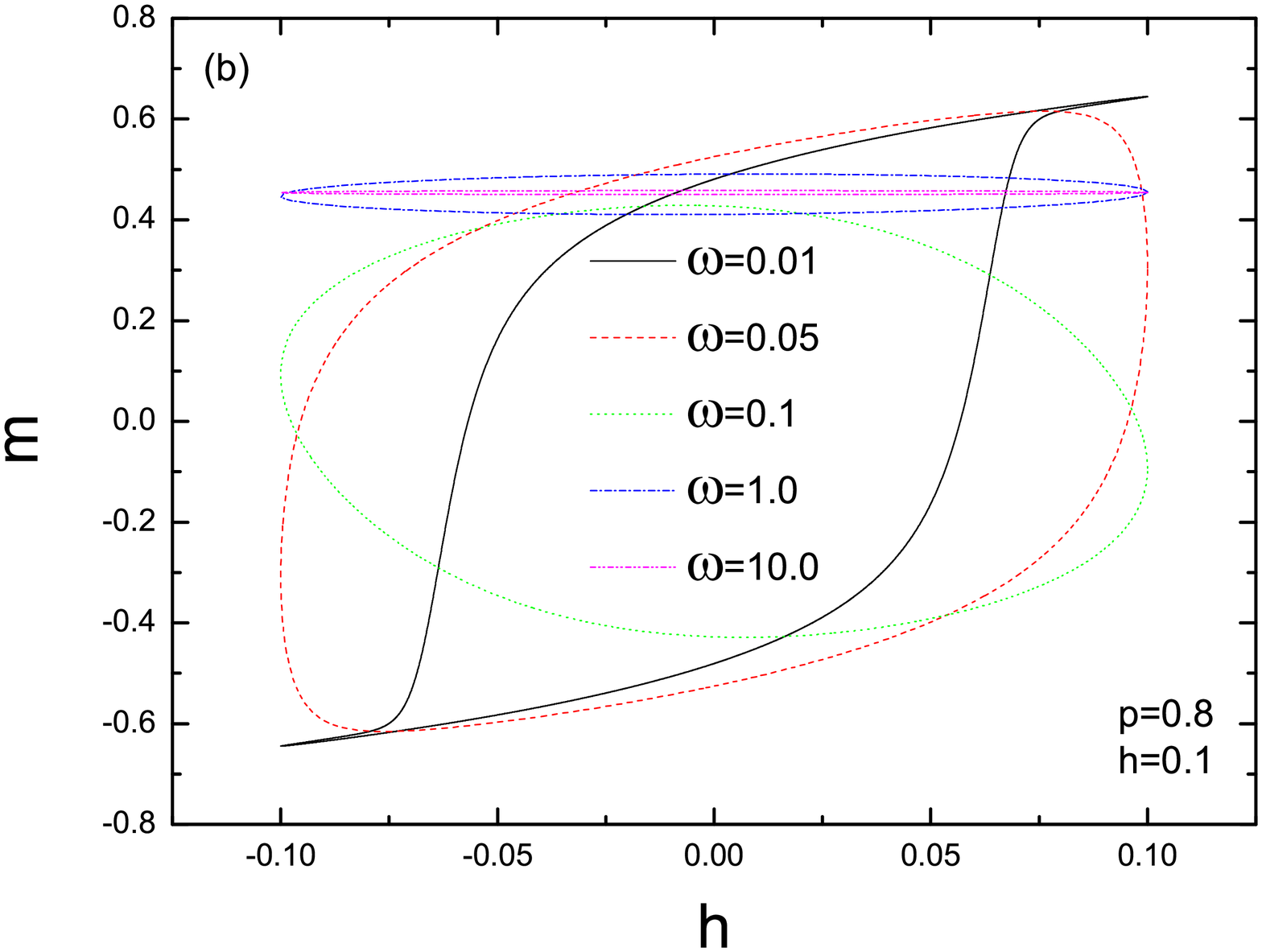}
\includegraphics[width=6.0cm]{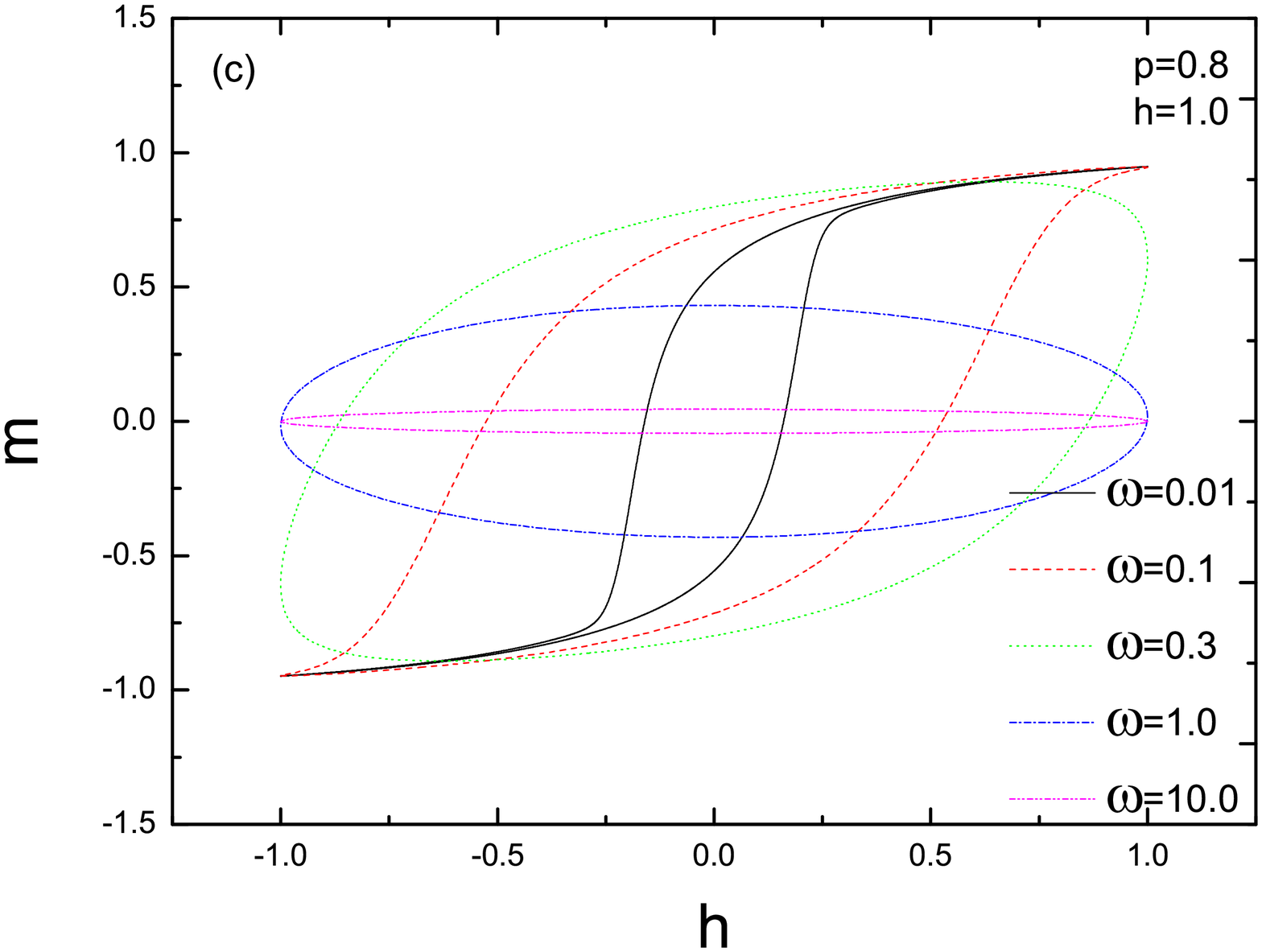}\\
\caption{Hysteresis loops of the quenched bond diluted kinetic Ising system for several values of applied field frequencies and
amplitudes with some selected values of active bond concentrations.}\label{fig6}
\end{figure}

The two fundamental tools, namely coercivity and remanence properties allow us to make predictions on the
shape of the hysteresis loops. Effect of the field frequency on the shape of typical hysteresis loops
corresponding to Figs. \ref{fig3}, \ref{fig4}, and \ref{fig5} are depicted in Fig. \ref{fig6} for a wide variety of
oscillation frequencies within the range $0.01\leq\omega\leq10.0$ with given values of $p$ and $h$. In Fig. \ref{fig6}(a),
we present the hysteresis loop curves corresponding to type-I characteristics of DLA curves. As seen in Fig. \ref{fig6}(a),
hysteresis curves exhibit positive valued narrow loops at low frequencies then evolve into a horizontal line with increasing
frequencies. Type-II curves are shown in Fig. \ref{fig6}(b), where the low frequency paramagnetic loops turn into high frequency
ferromagnetic loops due to a frequency induced dynamic symmetry loss in the system. Moreover, according to Fig. \ref{fig6}(c),
symmetric type-III hysteresis curves keep their dynamic paramagnetic state and exhibit remanence and coercivity with
branches having both positive and negative values. We can conclude from Fig. \ref{fig6}(c) that low frequency
thin and $S$-shaped hysteresis loops evolves into broad square-like shapes then turn into elliptical loops for sufficiently
high frequencies. If we increase the frequency further, these elliptical loops become horizontal lines.
Related to the frequency evolutions of the hysteresis curves, we should also note that a qualitative consistency is
found  between our numerical findings and experimental observations \cite{he, jiang, liu2}.

\begin{figure}
\includegraphics[width=6.0cm]{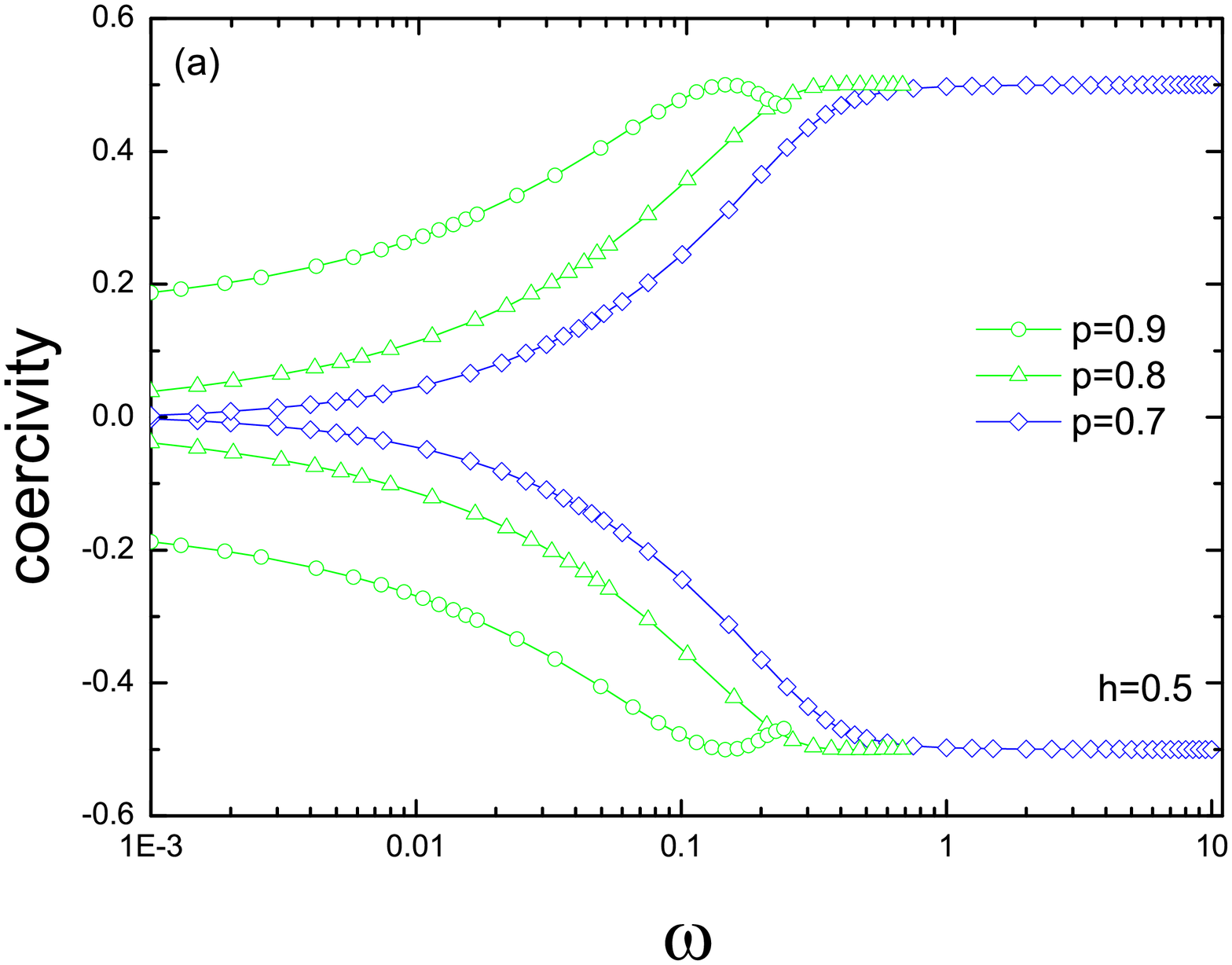}
\includegraphics[width=6.0cm]{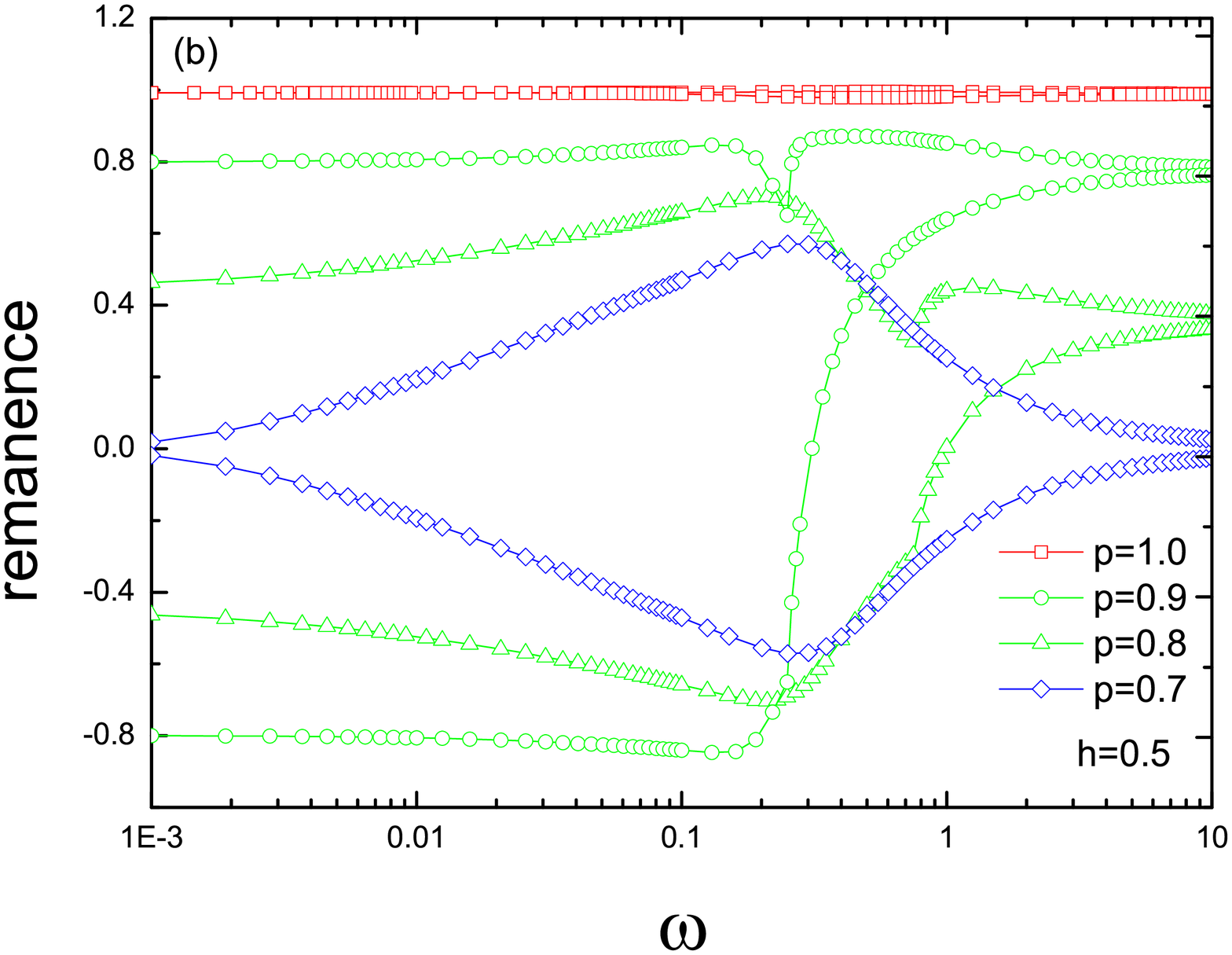}\\
\caption{Variation of  the coercive field (a) and  the remanent magnetization (b) as a function of the
applied field frequency for a field amplitude value of $h=0.5$ with some selected values of
disorder parameter $p$.}\label{fig7}
\end{figure}

In the following analysis, let us focus on the influence of disorder on the frequency dispersion of
hysteretic properties of the system. In Figs. \ref{fig7}(a) and \ref{fig7}(b), we represent variation
of coercivity and remanence curves as a function of oscillation frequency for some selected
values of disorder parameter $p$ with a given field amplitude $h=0.5$. From Fig. \ref{fig7}(a),
we see that as the concentration of diluted bonds increases then the zero frequency branches of
coercivity curves vanish as $\omega\rightarrow0$. In type-II curves, critical frequency at
which the system exhibit a dynamic phase transition increases with increasing disorder.
This frequency value can be regarded as a cut-off frequency at which the symmetric loops turn
into asymmetric loops. Hence, the non-symmetric coercivity data has not been considered in
Fig. \ref{fig7}(a). Moreover, coercivity curves exhibiting type-II characteristics may evolve
into type-III behavior, if the concentration of diluted bonds exceeds a critical
value (i.e. percolation threshold). Hence, we understand that the system exhibits a disorder
induced dynamic phase transition. In Fig. \ref{fig7}(b), we illustrate the dependence of
frequency dispersion of remanence curves on the concentration of disordered bonds with
a fixed external field amplitude $h=0.5$ corresponding to Fig. \ref{fig7}(a).
Type-I curves in Fig. \ref{fig7}(b) exhibit a positive remanence which indicates that
the system stays in a dynamic ferromagnetic order for the whole range of frequency spectrum.
By increasing disorder, type-I curves turn into type-II characteristics where we can observe that
either branches of remanence curves tend to lose their symmetric shape with increasing frequency
which signals the appearance of a frequency induced dynamic phase transition. In a similar manner,
we observe in Fig. \ref{fig7}(b) that for sufficiently low concentration of magnetic bonds,
the remanent magnetization curves may exhibit type-III characteristics where the system always exhibit
a dynamically disordered phase for the whole range of oscillation frequency spectrum.

\begin{figure}
\includegraphics[width=6.0cm]{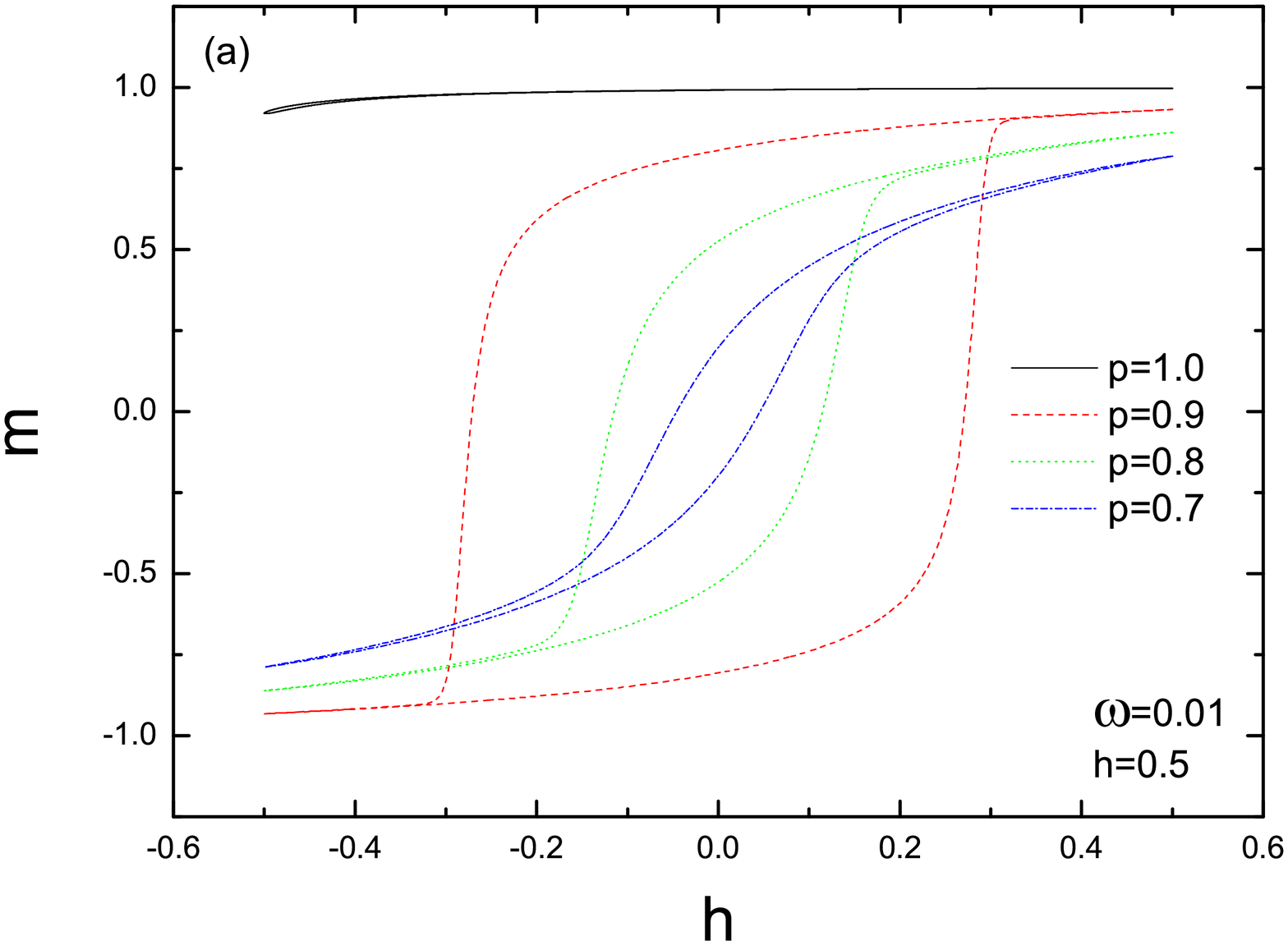}
\includegraphics[width=6.0cm]{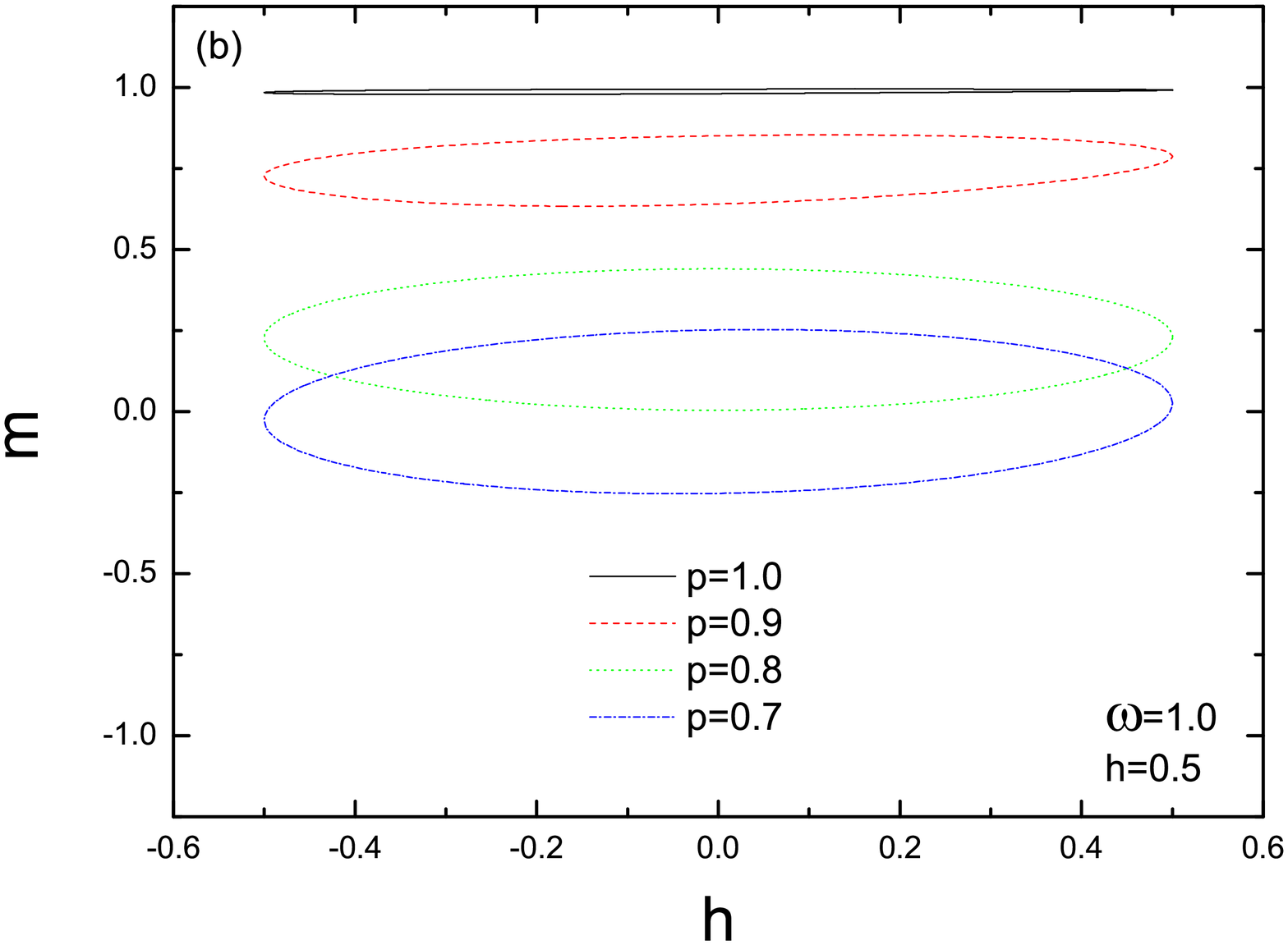}\\
\caption{Hysteresis curves of the quenched bond diluted kinetic Ising system for several values of active bond
concentrations  with  selected values of applied field frequencies and amplitude.}\label{fig8}
\end{figure}

As a final investigation, we discuss the evolution of some typical hysteresis loops
corresponding to remanence and coercivity properties depicted in Fig. \ref{fig7}.
In Fig. \ref{fig8} we fix the amplitude of the external magnetic field
as $h=0.5$. Fig. \ref{fig8}(a) corresponds to low frequency situation $(\omega=0.01)$.
From Fig. \ref{fig8}a, we see that a disorder induced dynamic phase transition
takes place in the system with increasing disorder. Hysteresis loops exhibit a
positive valued horizontal line in the pure case whereas as the amount of disorder
increases then the loop area gets narrower and we observe S-shaped thin loops.
On the other hand, the situation is depicted for relatively high frequency values
such as $\omega=1.0$ in Fig. \ref{fig8}(b). In this case, asymmetric loops corresponding
to ferromagnetic order become elliptical symmetric loops in the dynamic paramagnetic
phase with increasing disorder. Hence, we can conclude that influence of disorder on the
shape of hysteresis loops strictly depends on the magnitude of the applied field frequency.
Aforementioned results are also consistent with the observations illustrated in Fig. \ref{fig7}.

\section{Conclusions}\label{conclude}
In conclusion, by using EFT with correlations, we have investigated frequency dispersion features of
dynamic loop area, coercivity and remanent magnetization of a kinetic Ising model in the presence of
quenched-diluted bonds. First of all, in order to gain some general insight regarding the influence
of bond disorder, external field amplitude, and oscillation frequency on the dynamic phase transition
properties of the system we have demonstrated the phase diagrams of our model in a $(k_{B}T_{c}/J-\omega)$
plane for weak $(h=0.25)$, intermediate $(h=0.5, 0.75)$  and large $(h=1.0)$ values of the external field
amplitude with some selected values of disorder parameter $p$. We have found that critical frontiers
between symmetric and asymmetric phases in $(k_{B}T_{c}/J-\omega)$ plane are strongly dependent on
amplitude and frequency of oscillating field, as well as amount of disorder.

EFT is superior to conventional MFT predictions owing to the fact that the single spin correlations,
in other words thermal fluctuations are partially considered. Hence, we take the recently published
work performed by Ref.\cite{punya} one step forward by considering single-site correlations between
different lattice sites, and by considering the presence of disordered bonds. However, it is worth to
note that the approximation method used in the present work, does not take into account the local
structure of the lattice. In other words, EFT cannot distinguish between a two dimensional triangular
lattice and a three dimensional simple cubic lattice, since both have a coordination number $q=6$.
In this context, the results obtained in the present paper depend on the coordination number of the system.

The results obtained in the present paper qualitatively support the recent findings based on MFT \cite{punya}
for the pure case. Namely, we have observed that frequency dispersion of dynamic loop area (DLA) can be
classified into three distinct types. In type-I behavior, the system always exhibits asymmetric hysteresis
loops at any oscillation frequency for small field amplitudes and weak amount of disorder whereas type-II curves
are characterized by a frequency induced phase transition where the low frequency symmetric loops become
asymmetric loops with increasing oscillation frequency. On the other hand, type-III behavior is essentially
represented by symmetric hysteresis loops for the whole range of frequency spectrum when the external
field amplitude is sufficiently large and (or) if there exists sufficiently strong dilution of ferromagnetic
nearest-neighbor bonds. Therefore we can conclude that type-II curves exhibit frequency induced dynamic phase
transitions whereas evolution of DLA curves from type-I to type-III characteristics is due to a disorder
induced effect or as a consequence of large amplitudes of the external magnetic field.

Aforementioned observations have also been supported by detailed investigation of frequency dispersions of
coercivity and remanent magnetizations. From the technological point of view, in magnetic information storage
devices a material with high coercivity and large remanence is desired to preserve the stored data from unwanted
external factors. In this regard we have found that there exists an optimum field frequency for a disordered
material under an oscillating magnetic field at which the type-III remanence (i.e. the distance between positive
and negative remanent magnetizations corresponding to type-III) exhibit a maximum value, and as the field amplitude
increases then the optimum frequency value shifts towards higher frequency values. Finally, we have observed that
influence of disorder on the shape of hysteresis loops strictly depends on the magnitude of the applied field
frequency.

As a conclusion, we hope that the results obtained in this work would shed light on the further
investigations of the dynamic nature of the critical phenomena in disordered systems and would be
beneficial from both theoretical and experimental points of view.

\section*{Acknowledgements}
The authors (E.V. and Y.Y.) would like to thank the Scientific and Technological Research Council of Turkey (T\"{U}B\.{I}TAK) for
partial financial support. The numerical calculations reported in this paper were performed at T\"{U}B\.{I}TAK ULAKBIM (Turkish agency),
High Performance and Grid Computing Center (TRUBA Resources).



\begin{thebibliography}{99}
\bibitem{chakrabarti1_acharyya1} B.K.  Chakrabarti, and M. Acharyya, Rev. Mod. Phys. \textbf{71}, 847 (1999), and the references therein.
\bibitem{steinmetz} C.P. Steinmetz, Trans. Am. Inst. Electr. Eng. \textbf{9}, 3 (1892).
\bibitem{lo}  W.S. Lo, and R.A.  Pelcovits,  Phys. Rev. A \textbf{42}, 7471 (1990).
\bibitem{acharyya2_chakrabarti2} M. Acharyya,  and B.K. Chakrabarti, Phys. Rev. B \textbf{52}, 6550 (1995).
\bibitem{acharyya3} M. Acharyya, Physica A \textbf{253}, 199 (1998).
\bibitem{acharyya4} M. Acharyya, Phys. Rev. E \textbf{58}, 179 (1998).
\bibitem{rao} M. Rao, H.R. Krishnamurthy, and R. Pandit, Phys. Rev. B \textbf{42}, 856 (1990).
\bibitem{sides1} S.W. Sides, P.A.  Rikvold, and M.A. Novotny, Phys. Rev. E \textbf{59}, 2710 (1999).
\bibitem{sides2} S.W. Sides, P.A. Rikvold, and M.A. Novotny, Phys. Rev. Lett. \textbf{81},  834 (1998).
\bibitem{sides3} S.W. Sides, P.A. Rikvold, and M.A. Novotny, Phys. Rev. E  \textbf{57},  6512 (1998).
\bibitem{korniss} G. Korniss, P.A. Rikvold, and M.A. Novotny, Phys. Rev. E \textbf{66},  056127 (2002).
\bibitem{zhu} H. Zhu, S. Dong, and J.-M.  Liu,  Phys. Rev. B \textbf{70}, 132403 (2004).
\bibitem{zhong} F. Zhong, Phys. Rev. B \textbf{66}, 060401(R) (2002).
\bibitem{acharyya6} M. Acharyya, Phys. Rev. E \textbf{59},  218 (1999).
\bibitem{shi} X. Shi, G. Wei, and  L. Li,  Phys. Lett. A \textbf{372}, l5922 (2008).
\bibitem{deviren} B. Deviren,  O. Canko, and M. Keskin, Chin. Phys. B \textbf{19}, 050518 (2010).
\bibitem{tome} T. Tom\`{e}, and M.J. de Oliveira, Phys. Rev. A \textbf{41}, 4251 (1990).
\bibitem{acharyya5} M. Acharyya, J. Phys. A: Math. Gen. \textbf{27}, 1533 (1994).
\bibitem{punya} A. Punya, R. Yimnirun, P.  Laoratanakul, and Y. Laosiritaworn,  Physica B \textbf{405}, 3488 (2010).
\bibitem{lyuksyutov} I.F. Lyuksyutov, T. Nattermann, and V. Pokrovsky, Phys. Rev. B \textbf{59},  4260 (1999).
\bibitem{nattermann1} T. Nattermann, V. Pokrovsky, and  V.M. Vinokur, Phys. Rev. Lett. \textbf{87},  197005 (2001).
\bibitem{nattermann2} T. Nattermann, and V. Pokrovsky, Physica A \textbf{340},  625 (2004).
\bibitem{misra} A. Misra, and K. Chakrabarti, Europhys. Lett. \textbf{52}, 311 (2000).
\bibitem{sch} F. Sch\"{u}tze, and  T. Nattermann, Phys. Rev. B, \textbf{83} 024412 (2011).
\bibitem{wang} L. Wang, B.H.  Teng, Y.H. Rong, Y. L\"{u}, and Z.C.  Wang, Sol. State Comm. (2012) http://dx.doi.org/10.1016/j.ssc.2012.04.076
\bibitem{zimmer} M.F. Zimmer, Phys. Rev. E \textbf{47},  3950 (1993).
\bibitem{luse} C.N. Luse, A. Zangwill, Phys. Rev. E \textbf{50}, 224 (1994).
\bibitem{liu} J.-M. Liu, H.L.W. Chan, C.L. Choy, and C.K. Ong, Phys. Rev. B \textbf{65}, 014416 (2001).
\bibitem{sariyer} O.S. Sar{\i}yer, A. Kabakcioglu, and A.N. Berker, arXiv.1206.0230v1
\bibitem{mayergoyz} I.D. Mayergoyz, Mathematical Model of Hysteresis, Springer-Verlag, Berlin, (1991).
\bibitem{gilmore} R. Gilmore, Phys. Rev. A \textbf{20},  2510 (1979).
\bibitem{jiang} Q. Jiang, H.-N. Yang, and G.-C. Wang, Phys. Rev. B \textbf{52}, 14911 (1995).
\bibitem{robb} D.T. Robb, Y.H. Xu,  O. Hellwig, J.  McCord, A.  Berger, M.A. Novotny, and P.A. Rikvold, Phys. Rev. B \textbf{78}, 134422 (2008).
\bibitem{choi} B.C. Choi, W.Y. Lee, A. Samad, and J.A.C. Bland Phys. Rev. B \textbf{60}, 11906 (1999).
\bibitem{he} Y.-L. He, and G.-C. Wang, Phys. Rev. Lett. \textbf{70}, 2336 (1993).
\bibitem{lee} W.Y. Lee, B.-Ch. Choi, J. Lee, C.C. Yao, Y.B. Xu, D.G. Hasko, and J.A.C. Bland, Appl. Phys. Lett. \textbf{74}, 1609 (1999).
\bibitem{lee2} W.Y. Lee, B.-Ch. Choi, Y. B. Xu, and J.A.C. Bland, Phys. Rev. B \textbf{60},  10216 (1999).
\bibitem{rivera} A.R.-Rivera, J.M. Ferreira, and F.C. Montenegro, J. Magn. Magn. Mat. \textbf{226-230}, 1309 (2001).
\bibitem{santi}  L. Santi, R.L. Sommer, A. Magni, G. Durin, F. Colaiori, and S. Zapperi, IEEE Trans. Magn. \textbf{39}, 2666 (2003).
\bibitem{moore} T.A. Moore, J. Rothman, Y.B. Xu, and J.A.C. Blanda, J. Appl. Phys. \textbf{89}, 7018 (2001).
\bibitem{zheng_li} G.-P. Zheng, and  M. Li, Phys. Rev. B \textbf{66}, 054406 (2002).
\bibitem{akinci} U. Akinci, Y. Y\"{u}ksel, E. Vatansever, and H. Polat, Physica A (2012), doi: 10.1016/j.physa.2012.06.060.
\bibitem{vatansever} E. Vatansever, B.O. Aktas, Y. Y\"{u}ksel, U. Akinci, and H. Polat,  J. Stat. Phys. \textbf{147}, 1068 (2012).
\bibitem{glauber} R.J. Glauber, J. Math. Phys. \textbf{4}, 294 (1963).
\bibitem{honmura_kaneyoshi} R. Honmura, and T. Kaneyoshi,  J. Phys. C \textbf{12}, 3979 (1979).
\bibitem{kaneyoshi1} T. Kaneyoshi, Acta Phys. Pol. A \textbf{83}, 703 (1993).
\bibitem{tamura_kaneyoshi} I. Tamura, and T. Kaneyoshi,  Prog. Theor. Phys. \textbf{66}, 1892 (1981).
\bibitem{huang} Z. Huang, Z. Chen, F. Zhang, and Y. Du,  Phys. Lett. A \textbf{338}, 485 (2005).
\bibitem{altavilla} C. Altavilla, E. Ciliberto, Inorganic Nanoparticles: Synthesis, Applicaions, and Prespectives, Taylor\&Francis, CRC Press, p33 (2010).
\bibitem{liu2} J.-M. Liu, Q. Xiao, Z.G. Liu, H.L.W. Chan, and N.B. Ming,  Mat. Chem. Phys. \textbf{82}, 733 (2003).
\end{thebibliography}
\end{document}